\documentclass{aa}
\usepackage{amssymb}
\usepackage{graphicx}
\usepackage{hyperref}

\def\asec{\ifmmode ^{\prime\prime}\else$^{\prime\prime}$\fi}

\def\msun{M$_{\odot}$}

\def\asec{\ifmmode ^{\prime\prime}\else$^{\prime\prime}$\fi}

\def\ks{$K_{\rm s}$}
\def\degs{\ifmmode ^{\circ}\else$^{\circ}$\fi}
\def\amin{\ifmmode ^{\prime}\else$^{\prime}$\fi}
\def\farcm{\hbox{$.\mkern-4mu^\prime$}}

\begin{document}

\title{The late-time 
light curve of the Type Ia supernova 2000cx
\thanks{Based  on observations collected at the European Southern Observatory,
Paranal, Chile (ESO Programmes 67.D-0134 and 68.D-0114).
Also based in part on observations with the NASA/ESA {\it Hubble
Space Telescope}. 
These {\it HST}
observations are associated with proposals GO-8602 and GO-9114.}}

\author{
Jesper~Sollerman\inst{1} 
\and
Jan~Lindahl\inst{1,2} 
\and 
Cecilia~Kozma\inst{1}
\and
Peter Challis\inst{3}
\and
Alexei~V.~Filippenko\inst{4}
\and
Claes Fransson\inst{1}
\and
Peter M. Garnavich\inst{5}
\and
Bruno Leibundgut\inst{6}
\and
Weidong Li\inst{4}
\and
Peter Lundqvist\inst{1}
\and
Peter Milne\inst{7}
\and
Jason Spyromilio\inst{6}
\and
Robert P. Kirshner\inst{3}
}

\institute{
Stockholm Observatory, AlbaNova, Department of Astronomy, 106 91 Stockholm, Sweden 
\and 
Onsala Space Observatory, Chalmers University of Technology, 439 92 Onsala, Sweden
\and
Harvard-Smithsonian Center for Astrophysics, 60 Garden Street, Cambridge, MA 02138, USA
\and
Department of Astronomy, University of California, Berkeley, CA  94720-3411, USA
\and
Physics Department, University of Notre Dame, Notre Dame, IN 46566, USA
\and
European Southern Observatory, Karl-Schwarzschild-Strasse 2,
D-85748 Garching, Germany 
\and
Steward Observatory, University of Arizona, 933 North Cherry Avenue, Tucson, AZ 85721, USA
}

\date{Received --- ?, Accepted --- ?} 
\authorrunning{Sollerman et al.}
\titlerunning{SN 2000cx @ late phases} 
\offprints{jesper@astro.su.se}

\abstract
{We have conducted a systematic and comprehensive monitoring programme
of the Type Ia supernova 2000cx at late phases using the VLT and
{\it HST}. The VLT observations cover phases 360 to 480 days past maximum
brightness and include photometry in the $BVRIJH$ bands, together with a 
single epoch in each of $U$ and \ks.  While the optical
bands decay by about 1.4 mag per 100 days, we find that the
near-IR magnitudes stay virtually constant during the observed
period. This means that the importance of the near-IR to the
bolometric light curve increases with time. The finding is also in
agreement with our detailed modeling of a Type Ia supernova in the
nebular phase. In these models, the increased importance of the
near-IR is a temperature effect.  We note that this complicates
late-time studies where often only the $V$ band is well monitored. In
particular, it is not correct to assume that any optical band follows
the bolometric light curve at these phases, and any conclusions based
on such assumptions, e.g., regarding positron-escape, 
must be regarded as premature.
A very simple model where all positrons are trapped
can reasonably well account for the 
observations. The nickel
mass deduced from the positron tail of this light curve is
lower than found from the  peak brightness, providing an estimate of the 
fraction of late-time emission that is 
outside of the observed wavelength range.
Our detailed models show the signature of an infrared catastrophe at these 
epochs, which is not supported by the observations. 
}

\maketitle

\section{Introduction}

Type Ia supernovae (SNe~Ia) are believed to be the destructive thermonuclear
explosions of white dwarfs 
(e.g., Hoyle \& Fowler 1960; see for a review Leibundgut 2000). 
During the first years, their light curves are 
powered by the
radioactive decay of freshly nucleosynthesized $^{56}$Ni, releasing
MeV $\gamma$-rays and positrons in the ejecta. The uniformity of the 
early-time  light curves has enabled the use of SNe~Ia for 
measuring cosmological distances
(Schmidt et al. 1998; Riess et al. 1998; Perlmutter et al. 1999). 

The peak luminosity of such an explosion is given by the amount of
ejected $^{56}$Ni, and the 
early bolometric light curve can thus be used to determine this important 
quantity
(Arnett 1982; Contardo et al. 2000).
The shape of the light curve at the early post-maximum phase 
(40--120 days past explosion) is
determined by the fraction of the $\gamma$-rays that are thermalized.
While the massive Type II SNe (like SN 1987A) generally trap
most $\gamma$-rays, resulting in late-time light curves tracking the
radioactive decay of 0.98 mag per 100 days, the light curves of SNe Ia
decline much faster. This is because the small progenitor masses and
the high expansion velocities of SNe Ia 
quickly make the ejecta transparent to the $\gamma$-rays. 

At phases later than about 200 days past the explosion, virtually all
$\gamma$-rays escape freely from the ejecta, and the luminosity is
then provided by the kinetic energy deposited by the positrons. 
Whether or not the positrons are able to escape from
the ejecta depends on the strength and geometry of the magnetic field
(Colgate et al. 1980; Ruiz-Lapuente \& Spruit 1998; 
Milne et al. 1999, 2001). While a
weak and radially combed magnetic field might allow 
an increasing fraction of the positrons to escape,
thus providing a steep light curve, a strong, tangled magnetic field
would efficiently trap all the positrons, and keep the light curve at
the radioactive decay rate. 

The study of the positron phase of the light curve could thus potentially 
help answer questions about the magnetic field configuration in exploding 
white dwarfs, as well as to constrain the contribution of supernova explosions
to the diffuse Galactic 511 keV background.
Previous attempts to study the positron phase of the light curve 
have, however, been hampered by the lack of both observational data and
detailed modeling.

As a first step to investigate whether observations of
the positron phase can establish any conclusions about positron escape, we 
have attempted a photometric study at late phases of the
Type Ia SN 2000cx, and modeled its light curve in some detail.

\subsection{SN 2000cx}

SN 2000cx was discovered on 17.5 July 2000 (Yu et al. 2000; UT dates 
are used throughout our paper). It was located $23\farcs0$ west and
$109\farcs3$ south of the nucleus of the S0 galaxy NGC 524. It was
classified as a SN~Ia, with a spectrum resembling that of
the peculiar SN 1991T 
(Chornock et al. 2000; see Filippenko 1997 for a review of supernova
spectra).  The recession velocity of NGC 524 is only
$\sim2400$~km~s$^{-1}$, and SN 2000cx became the brightest supernova
observed that year. Furthermore, the favorable position far from the
galaxy center made it a good target for late-time photometry. The early-time 
photometry has been extremely well covered by Li et al. (2001)
and by Candia et al. (2003). The $B$-band maximum occurred on July 26.3 
(MJD 51752.2) at $B = 13.42$ mag. The early decline rate was 
determined to be
$\Delta m_{15}(B) = 0.93$ (Li et al. 2001). 
The light curve around peak was rather unusual.
It was a fast riser but a slow decliner (Li et al. 2001),
and SN 2000cx has 
therefore been labeled a peculiar supernova. 
The unusual early light curve may have
been related to the presence of high-velocity material (Branch et al. 2004).

Here we report on our late-time observations of SN 2000cx
obtained with the Very Large Telescope (VLT) and the {\it Hubble
Space Telescope (HST)}. Preliminary results have been reported by
Sollerman et al. (2004) and the results presented here 
supersede that report.
The observations, data reduction,
and photometry are described in Sect. 2. 
The results are presented in Sect. 3, 
and in Sect. 4 
we discuss the late-time light curves of SNe~Ia in general. 
In Sect. 5 
we describe our detailed modeling, and Sect. 6 discusses the
modeling in conjunction with the data.
Our main conclusions are summarized in Sect. 7.\\
\vspace{1.0cm}

\section{Observations and data analysis}
\subsection{Optical imaging with the VLT}

The field of SN 2000cx was observed in the optical ($UBVRI$) 
range during several epochs between 359 and 480 days past maximum light.
These observations were obtained with the Focal Reducer and 
low dispersion Spectrograph (FORS) 
instruments\footnote{http://www.eso.org/instruments/fors1/} at the 
European Southern Observatory (ESO) VLT. 
FORS2 was used for the first epoch of observations. 
This instrument was attached to 
the fourth of the four VLT Unit 
telescopes (UT4).  The subsequent observations were all conducted 
with FORS1, which was then mounted at the Cassegrain focus of 
UT3. 
A log of the observations is given in Table~1.

Both of the FORS instruments
were equipped with $2048 \times 2048$ pixel CCDs having $24~\mu$m pixels. 
We used the standard resolution mode that gives a field-of-view of 
$6\farcm8 \times 6\farcm8$ with a pixel size on the sky of 0\farcs20.
An image of the field of the supernova is displayed in Fig.~1, and
a detail of the region around the supernova is shown in Fig.~2.

All observations were split into several exposures with small telescope 
shifts in between. This was to avoid heavy saturation of a nearby 
bright star, and to allow for removal of cosmic rays. The observations 
were all conducted in service mode by the Paranal staff.

The data were reduced with
the NOAO {\tt IRAF} software.\footnote{IRAF, http://iraf.noao.edu/iraf/web/,
is distributed by NOAO.}
We have bias-subtracted and twilight-sky flattened all 
the frames and then combined the exposures in the same filters using
a sigma clipping rejection.

Our first and last epochs of observation 
were photometric and on the first epoch we observed the 
following standard fields of Landolt (1992) with FORS2: Mark~A, 
PG~1323-086, PG~2331+055, and SA~109-949. There were in total 15 stars
useful for obtaining the photometric solutions in these fields. 
The instrumental magnitudes of the standard stars
were all measured with an aperture of 15 pixel radius. The sky background
was estimated using an annulus of inner radius 30 pixels and outer radius 
40 pixels. We used the package {\tt PHOTCAL} to calculate the transformation
equations to determine zero points, color terms, and atmospheric 
extinction terms.  

The color terms we obtained were actually very similar 
to the averaged transformations provided by
ESO.\footnote{www.eso.org/observing/dfo/quality/FORS2/qc/photcoeff/photcoeffs$\_$fors2.html}
Since the ESO measurements are based on a larger number of standard stars,
we decided to fix the color and extinction terms to the values determined by 
ESO.

This procedure allowed us to calibrate a number of local stars 
in the field of SN 2000cx. 
Photometry of the supernova was then tied to the system of Landolt 
standards via these field stars, so that accurate supernova photometry 
could be established also on non-photometric nights.

To check the nightly zero points we also reduced the standard
stars for the last epoch of observations. On that night we observed 3
standard fields with FORS1: PG2213-006, Rubin 152, and SA100-269, with
a total of 9 useful standard stars (Landolt 1992). 
For the 14 local standard stars chosen all over the CCD
frame, the magnitude differences between the first and last epochs had 
a root-mean-square (rms) of 0.015 mag in the $V$ band. The largest difference, in the 
$B$ band, 
was still less than 0.04 mag. The magnitudes for our local standards are
given in Table~2, together with their offsets from the supernova. 
These magnitudes were measured in a 6-pixel radius
aperture, and aperture corrected using the task {\tt MKAP}. 
Errors in the aperture correction were typically less than 0.01 mag, 
but reached 0.016 mag in the $U$ band. 
The magnitude errors for the standard stars reported in
Table~2 are the photometric errors from {\tt PHOT} added in quadrature to 
the rms difference in magnitude between the two epochs.
The error budget for the local standards is certainly below 0.05 mag
in all bands, possibly with the exception of the $U$ band, which is always more
difficult (e.g., Suntzeff 2000). 
The $U$-band uncertainties in Table~2 are simply the error
estimates from {\tt PHOT}, including errors for the transformations.


The supernova photometry was performed on each epoch in a 6-pixel
radius aperture.  The magnitudes for the supernova are given in
Table~3.   These magnitudes are aperture-corrected and 
tied to the Landolt standards via the
local standards.  Here we have estimated the supernova magnitudes as
the mean value of the differential magnitudes compared to our 14 local
standards as given in Table~2.  The estimated uncertainties given in Table~3
are combined from the error estimate by {\tt PHOT} and the
standard deviation for the 14 different magnitude estimates for each
filter and epoch. The latter error should to some extent also encapsulate
uncertainties in aperture corrections, color transformations, and
zero points. 

\subsection{Infrared observations}

Near-infrared observations were obtained using the 
Infrared Spectrometer And Array 
Camera (ISAAC)\footnote{http://www.eso.org/instruments/isaac/} 
attached to UT1 of the ESO VLT.
A log of the observations is given in Table~4, and regions of 
two images are displayed in Fig.~2.

We used the Short Wavelength (SW) imaging mode with the
Rockwell Hawaii  HgCdTe $1024 \times 1024$ pixel array detector.
The pixel size on the sky was 0\farcs147 and the field-of-view 
was 2\farcm5 $\times$ 2\farcm5.
The observations were performed in the 
$J$, $H$, and \ks\ bands in jitter mode, with small offsets performed between 
each image in order to allow for sky subtraction. 
As the supernova was positioned far out from the core of the galaxy, 
we did not have to dither separately for sky.

For the $J$ band, the detector integration times (DITs) were 30~s, and
each observation was built up from 3 DITs per exposure (NDITs).
This means that we stayed on the same spot in the sky for $\sim1.5$ minutes.
The number of exposures (NEXP) in each observational block 
was typically 15 in the $J$ band, which gives total exposure times 
(NDIT$\times$DIT$\times$NEXP) of  1350~s per observing block in the $J$ band.
The exposure times for the other bands can be seen in Table~4. 
We performed \ks\ observations of SN 2000cx only for the first epoch, due to 
our limited amount of observing time.

The data were reduced with
the {\tt Eclipse}\footnote{http://www.eso.org/projects/aot/eclipse/}
and NOAO {\tt IRAF} software. 
Dark and flatfield images were prepared
using standard {\tt Eclipse} recipes.
The sky background level was subtracted and the images were summed 
using the routine {\tt jitter} 
within {\tt Eclipse}. 

Photometric standard stars are observed regularly with ISAAC in service mode.
We reduced the $J$, $H$, and \ks\ images of S279-F  (Persson et al. 1989) 
from our first epoch, as well as $J$ and $H$ images of S234-E 
(Persson et al. 1989), FS10, FS19, and FS32 (Hawarden et al. 2001) 
obtained on 
our last epoch of observations.
From these images we measured the flux of the standard stars to 
establish zero points for the given night and passband.
The average Paranal atmospheric
extinction of  0.11 mag airmass$^{-1}$ was used in the $J$ band, as well as
0.06 and 0.07 mag airmass$^{-1}$ in the $H$ and \ks\ bands, 
respectively.\footnote{http://www.eso.org/instruments/isaac/imaging\_standards.html}

In this way we were able to establish local standard stars in the field 
of SN 2000cx, although due to the smaller field of view of ISAAC, 
these were not the same stars as in the optical.
The magnitudes of the local standard stars were
measured using aperture photometry, and the magnitudes were
aperture corrected using isolated and well-exposed stars.
The magnitudes for the standard stars are shown in Table~5. 
The listed errors in $J$ and $H$ are the errors from 
{\tt PHOT} and the rms magnitude difference between the two epochs added in 
quadrature. For the \ks\ band we have only listed errors from {\tt PHOT}, 
since this band was used only at the first epoch.  Thus,
the somewhat smaller errors listed in this band are likely to be
underestimated accordingly.


The magnitudes of the supernova were again measured using differential
photometry, and the uncertainties estimated as rms errors from comparison
with  the 14 local standard stars,  added in quadrature with the
errors from {\tt PHOT}.  The supernova near-IR magnitudes are listed
in Table~6.  We note that the supernova itself was fainter than most
of our local standard stars at these late phases.  This means that the
error budget estimated by comparison with the standard stars is
probably too optimistic.  At these faint levels, sky subtraction
errors in the near-IR are likely to dominate the errors.
In fact, we verified our reductions by an
independent estimate using the {\tt XDIMSUM} package.  However, by
measuring ten sources close to the supernova, which all had magnitudes
comparable to the supernova itself,  we found a significant scatter in
the estimated magnitudes between the different epochs. From this
exercise, we  instead estimate an error of $\pm 0.15$ mag in
the supernova near-IR magnitudes.

\subsection{Late HST photometry}

SN 2000cx was also observed with {\it HST}. A log of the 
{\it HST} observations is given in Table~7.
 
We retrieved the data from the {\it HST} science archive, which delivers
the fully reduced WFPC2 images.  The photometry was performed using
the {\tt HSTPhot} package as described by Dolphin (2000).  This
software fits point-spread functions (PSF) to the detected
objects,  and also accounts for charge transfer efficiencies and
aperture corrections.  Following Li et al. (2002), we used option 10
to measure the supernova magnitudes.  This option uses local sky
determination and default aperture corrections.  The magnitudes in the
WFPC system and error estimates from {\tt HSTPhot} for SN 2000cx are
also given in Table~7.  An image of the supernova in the PC chip was
presented by Li et al. (2002, their Fig.~1 bottom right).  
The supernova lies in a clean and isolated region, which ensures
that aperture photometry in our high quality 
ground-based data is adequate for this object. We also do not see any signs 
of a light echo 
(cf. Schmidt et al. 1994). 
No other star is detected 
close to the supernova position; moreover, the $V-K$ and near-IR 
colors are not consistent with any stellar type, ruling out significant
contamination from a foreground star in the late-time near-IR data.

\subsection{Optical spectroscopy}

At the first epoch of optical observations, 360 days past maximum,
we also performed spectroscopy of SN 2000cx. On 24 July 2001 we obtained
a 2400~s exposure using FORS2 on UT4. We used the 300V grism
together with order-sorting filter GG375 and a $1\farcs3$ wide
slit. Two nights later, on 26 July 2001, this was complemented by
another 2400~s spectrum with the 300I grism and the OG590
filter. Together these two spectra cover the wavelength region 
$\sim$ 3700--9200~\AA.

The spectra were reduced in a standard way, including bias
subtraction, flat fielding, and wavelength calibration using spectra of
a Helium-Argon lamp.  Flux calibration was done relative to the
spectrophotometric standard star  LTT377 with the 300V grism and to 
G158-100 with the 300I grism. 
The absolute flux calibration of the combined spectrum was
obtained relative to our broad-band photometry.

\section{Results}

In Fig.~3 we plot the $UBVRI$ photometry from our VLT observations. The flux
from the supernova can be  seen to  decrease exponentially. A linear
fit to the magnitudes gives the slopes presented in Table~8.  The $BVR$
light curves decline by about 1.4 mag per 100 days.  
This is consistent with the SNe Ia light curve slopes at earlier epochs 
(e.g., Turatto et al. 1990), and also 
reasonably consistent with the early light curve of SN~2000cx 
as published by Li
et al. (2001).  In Fig.~4 we plot the $V$ and $R$ light curves from
that study, together with our late-time data. The dashed lines are
the extrapolations of the late-time slopes to the early-time data.
The $I$ band, on the other hand, does not follow the other bands at late
phases.  At 0.88 mag per 100 days, this passband declines
significantly more slowly.

The color evolution between 360 and 480 days past maximum is not very strong
in the optical range.  In Fig.~5 we show the $B-V$, $V-R$, and $R-I$
color evolution. Only in the $R-I$ color can a clear trend be seen, due
to the shallower slope in the $I$-band light curve.

The near-IR magnitudes from the VLT are plotted in Fig.~6. There is no
evidence for any fading of the supernova flux during the observed epoch. 
This result is new and shows that substantial color evolution is indeed
present at these very late phases. A linear fit to the $JH$ light curves gives 
the decline rates presented in Table~8. The $1\sigma$ 
errors on the decline rate
were derived for a photometric error of $\pm0.15$ mag at each date.

Finally, also 
the very late-time observations with the {\it HST} have been added to
the  earlier light curves in Fig.~4. The earliest {\it HST} data, from day
348,   have already been published by Li et al. (2002). They measured
the  WFPC2 magnitudes in the very same way as we have done, and with
the same  results (their Table~1). The magnitude we have plotted in
the $R$-band light curve in Fig.~4 is the F675W magnitude, and should
be converted to the  Cousins $R$-band system. Li et al. (2002) obtained
$R = 22.14$ mag with a
rough conversion using the transformation from Dolphin 
(2000).\footnote{Li et al. (2002) report $R = 22.10$ mag from an
approximation, but the more detailed calculation for these transformations
gives $R=22.14$ mag.} 
This is still consistent with our late light curve.  In
fact, as also noted by Li et al. (2002), the transformations used for
converting magnitudes between these systems are not really applicable
to emission-line objects.  A better method may be to use our late-time
spectrum obtained 360 days past  maximum, to define the filter
transformation.  We used this spectrum, and  integrated the flux under
the filter functions for both the $R$-band  filter and the {\it HST} F675W
filter using the SYNPHOT task {\tt pltrans}.  This gave an offset
of 0.23 (Vega) mag, which converts our measured  F675W
magnitude to $R = 22.21$ mag, and similarly F814W to $I = 21.68$ mag. Both
transformations would make the {\it HST} magnitudes undershoot the light
curves from the VLT, as illustrated by the added error bar in Fig.~4
for the transformed $R$-band magnitude at day 348. The later F555W
magnitudes would accordingly have 0.15 mag added to convert to the $V$
band. This is, however, 
under the assumption that the spectrum stays constant,
and we have not taken this transformation 
into account in Fig.~4 or in the following discussion, where such an
offset would not affect our conclusions in any way.

\subsection{Bolometric luminosity}

The wide $UBVRIJHK$ coverage of our broad-band magnitudes allow an
attempt to construct a UV-optical-IR (hereafter UVOIR) ``bolometric'' 
light curve. 
Such a light curve should represent 
the fraction of 
$\gamma$-ray and positron energy that is thermalized
in the ejecta, although at these late
phases 
some fraction of the energy is also
escaping at even longer wavelengths.

To derive an absolute bolometric light curve we need the distance and
reddening to SN 2000cx. The host extinction is expected to be small, since
the supernova occurred in an S0 galaxy, and far outside the central
region. Early high-resolution spectra show no evidence for
host galaxy Na I D absorption (Lundqvist et al. 2005).  There
is, however, a weak Galactic component of Na~I~D, consistent with the
galactic reddening in the direction of SN 2000cx as derived by 
Schlegel et al. (1998), 
$E(B-V) = 0.082$ mag. We will use this value
for the total extinction toward SN 2000cx.
The distance to SN 2000cx was reviewed in Candia et al. (2003) and is
not very well 
constrained.\footnote{The distance measurements 
to NGC 524 are actually quite disaccording. 
Surface Brightness Fluctuation measurements
indicate a distance modulus of $31.84\pm0.20$, while a supernova estimate gives $32.53\pm0.35$.} 
We will use a distance modulus of 32.47 mag in accordance with their analysis.


We made a simple UVOIR light curve in the following way. 
For the earlier epochs, we use the optical data ($BVRI$) from Li et al. (2001),
the near-IR data ($JHK$) 
from Candia et al. (2003) and assume that $U \approx B + 0.2$
(Candia et al. 2003, their Fig.~3). 
For our late VLT observations we assume that the $U-B$ and $H-K$ colors 
are constant during these epochs.
We convert the magnitudes to flux using
Fugutaki et al. (1995) and Allen (2000) 
for the optical and near-IR magnitudes, respectively.
We then simply integrate the flux from $U$ (0.36~$\mu$m) to $K$ (2.2~$\mu$m) 
to form the UVOIR light curve.
No assumptions are made on the fraction of emission lost outside this band 
(but see Suntzeff 1996). Where observations are missing, we simply 
interpolate. The extinction correction is done using the extinction 
law from Fitzpatrick (1999). 
The resulting light curve is displayed in 
Fig.~\ref{bolometric}.

\section{Light curves of SNe Ia}

There exist a rich literature regarding the light curves of SNe Ia
(e.g., Leibundgut 2000, and references therein). Almost all such studies
concern the early phases of the supernova, when it is bright and
relatively easy to observe. The importance of these early phases has
been greatly emphasized with the realization that SNe~Ia can be used for
accurate measurements of cosmological distances.

The light curves of SNe~Ia are powered by the deposition in the
expanding SN ejecta  of the $\gamma$-rays and positrons produced by
the decay chain  $^{56}$Ni $\rightarrow ^{56}$Co $\rightarrow
^{56}$Fe.  The shape of the early-time light curve of a SN~Ia depends
essentially on the fact that optical photons created by the 
thermalization of the $\gamma$-rays and positrons do
not immediately escape from the  optically thick 
ejecta. 
Maximum light occurs when the instantaneous rates of deposition of hard
radiation  and emission of optical light are roughly equal (Arnett
1982). With time, the delay between energy deposition and
emission of optical radiation becomes progressively smaller.

In this work we instead concentrate on the late phases. In the
nebular phase, the ejecta are essentially optically thin.
The light curves at these phases are still driven by the
radioactive decay of $^{56}$Co, which decays on a timescale of 111.3
days.  However, the few 
SNe~Ia for which there exist
well-monitored light curves at phases up to several hundred days
decline substantially faster than this.
Actually, such observations were available even before the understanding
that $^{56}$Co powers the light curves, and early suggestions for
the powering included nuclei with faster decay times, such as
californium (Baade et al. 1956, see Colgate et al. 1997 for an
interesting historical perspective). 
From nucleosynthesis arguments it was soon realized 
that most of the burned material is iron-group
elements, and most importantly $^{56}$Ni.
The observed rapid decline rate of the light curves is now instead 
interpreted in terms of escape of the $\gamma$-rays from the ejecta. As
the ejecta expand, the optical depth decreases and a progressively larger
fraction of the $\gamma$-rays escapes thermalization.

A simple toy model for
the late emission can sometimes be instructive to understand the decay of a 
supernova light curve (e.g., Sollerman et al. 2002).
In this model the flux from the decay of $^{56}$Co evolves as
$\mathrm{e}^{-t/111.3} \times \left(1 - 0.966\mathrm{e}^{-\tau}\right)$,  
where the optical depth, $\tau = {(t_1/t)}^2$, 
decreases due to the homologous expansion, and
$t_1$ sets the time when the optical depth to $\gamma$-rays is
unity. Furthermore, 3.4$\%$ of the energy in these decays is in the
form of the kinetic energy of the positrons, which are here assumed to be
fully trapped.  Such a simplistic model 
can reasonably well fit the
'bolometric' light curve of a SN~Ia after the diffusion
phase, with only two free parameters 
(Fig.~\ref{toymodel}).

As the
$\gamma$-rays escape the ejecta, the relative importance of the positrons will
increase, and as soon as the trapping of the $\gamma$-rays decreases to
below 3.4\% the light curve enters the positron-dominated phase. The
shape of the bolometric light curve will then depend on whether or not
the positrons are fully trapped. The simple model described above
assumes full trapping, which is likely the case if the ejecta of SNe~Ia 
have a strong and tangled magnetic field. 
In this case the light curve should
flatten out and approach the decay time of $^{56}$Co in the 
positron-dominated phase (see Fig.~\ref{toymodel}).  
However, late-time observations of SNe~Ia in optical
passbands have indicated that the light curve continues to fall
rapidly also at epochs later than 200 days. 

This has been interpreted in two different ways. 
It could be that 
the positrons escape from the ejecta, and therefore cannot efficiently 
power the late-time light curve. Alternatively, more and more of the emission 
may emerge at longer wavelengths outside our observational coverage.
To investigate these scenarios was the main motivation for our observational 
campaign of SN 2000cx.

The progressive transparency of positrons 
was suggested by Arnett (1979) 
and this positron escape scenario was elaborated by Colgate et al. (1980). 
It was observationally 
investigated by Cappellaro et al. (1997) and more recently by 
Milne et al. (1999, 2001), who suggested that positron escape must 
indeed occur in at least some
SNe~Ia in order to explain their late-time light curves. Ruiz-Lapuente \&
Spruit (1998) reached similar conclusions and connected them to the
theory of the white dwarf magnetic field. If positron escape is indeed
needed to explain the rapidly declining late-time light curves, this
implies weak or radially combed magnetic fields. 

These comparisons between observations and simple models have, however, 
several caveats.  Most importantly, 
the observations of late-time light curves are relatively
sparse, and are available in a few passbands only.  It is therefore
not possible to construct true bolometric light curves. 
At earlier phases, attempts to construct UVOIR light curves for SNe Ia are 
indeed useful for comparisons to models 
(e.g., Contardo et al. 2000). At these phases, the UVOIR light 
curve embraces most of the emitted light 
(Suntzeff 1996; Contardo et al. 2000). 
This assumption is not necessarily valid at later phases, 
but has often been adopted due to the lack of detailed observations.

For example, the work of Cappellaro et al. (1997) simply assumed
that the late $V$-band light curve follows the bolometric light curve at
late phases. A similar assumption was made by Milne et
al. (1999, 2001). 
However, we note that as the 
input heating decreases and the ejecta expand, the temperature is
likely to decrease as well, and color evolution could then mimic the
effect of positron escape.

To address this question requires comprehensive modeling of the
nebular phase of the supernova. With a model that predicts where the
emission will emerge, direct comparisons with filter-band curves can be
made.  The attempts to build a self-consistent model for a nebular SN
Ia took off with the PhD thesis of Axelrod (1980).  He argued against
positron escape and instead suggested that the emission at late phases
would emerge in the far-IR, thereby introducing the concept of the 
``infrared catastrophe.'' This is a thermal instability that occurs when the
temperature drops below what is needed to excite the optical atomic
levels, and the cooling is abruptly taken over by the far-IR 
fine-structure lines.

To be able to interpret our observations of SN 2000cx, we have therefore
constructed detailed models for the late phases of SNe~Ia.

\section{Modeling}

We have modeled the emission from a SN~Ia in the nebular
phase,  100--1000 days after explosion. The adopted code is an updated 
version of the code described by Kozma \& Fransson (1998a), 
and applied to SN 1987A by Kozma \& Fransson (1998a, 1998b). 
Below is a brief summary of the model.

\subsection{The model}

The supernova ejecta are powered by the decay of radioactive isotopes
formed in the explosion. In the model we include the decays of
$^{56}$Ni,  $^{57}$Ni, and $^{44}$Ti. $^{56}$Ni first decays to
$^{56}$Co on a time scale 
of 8.8 days and then to stable $^{56}$Fe on a
time scale of 111.26 days (all decay times are e-folding). 
In the decay of $^{56}$Co a fraction 
(3.4\%) of the energy is released in the form of 
kinetic energy of the positrons.  
$^{57}$Ni decays rapidly (51.4~h) to $^{57}$Co, which then decays to stable 
$^{57}$Fe on a time scale of 391 days.
The decay time scale of $^{44}$Ti is 87 years 
and this decay chain also includes emission of
positrons.  At the epochs we are modeling, the decay of
$^{56}$Co dominates the energy input. The $\gamma$-rays emitted in the
decays scatter off free and bound electrons in the medium, and give
rise to fast electrons (with an energy of $\sim$0.01--1~MeV). These
nonthermal electrons, as well as the positrons emitted in the decays, 
deposit their energy by heating, ionizing, or
exciting the ejecta. The amount of energy going into these three
different channels depends on the composition and degree of
ionization and 
is calculated by solving the Boltzmann equation as
formulated by Spencer \& Fano (1954).  Details of this calculation are
given in Kozma \& Fransson (1992).  In our model the $\gamma$-ray and positron
deposition is calculated for the different compositions  and
ionizations. We assume that the positrons are
deposited locally, within the regions containing the newly
synthesized iron.

As input to our calculations we use the density structure, abundances,
and velocity structure from model W7 (Nomoto et al. 1984; Thielemann
et al. 1986). The model is spherically symmetric containing zones of
varying compositions. Also, the amounts of radioactive elements are taken from
the W7 input model. We have made no attempts to alter the model to accomplish
a better fit to the observations.

The temperature, ionization, and level
populations in each zone are calculated time-dependently. We find that
steady state is a good approximation up to $\sim500$ days. Thereafter,
time dependence becomes increasingly important.

In addition to the atomic data given in Kozma \& Fransson (1998a), the
code has been updated to model SNe~Ia. 
In particular we have extended and updated our treatment
of the iron-peak elements. We have included charge transfer
reactions between the iron ions 
(Liu et al. 1998)
and extended the ionization balance for cobalt.
For the iron ions we have updated the total
recombination rates and added rates to individual levels (Nahar 1996;
Nahar 1997; Nahar et al. 1997).
In addition to the iron ions we solve Co~II, Co~III, Ni~I, and
Ni~II as multilevel atoms.

\subsection{Line transfer}

For the line transfer we use the Sobolev approximation 
(Sobolev 1957, 1960; Castor 1970). 
The Sobolev approximation turns the line transfer
into a purely local process: either the photon is re-absorbed on the spot 
or it escapes the medium. 
As the ejecta are expanding homologously,
with a velocity much larger than the thermal velocity of 
the matter, this is a good approximation for an individual, well-separated 
line. 
However, especially in the UV, there are many
overlapping lines, and one can expect UV scattering to be
important. 
The effect of line scattering is
to alter the emergent UV spectrum, but it also affects the
UV field within the ejecta. The ionization of elements with low
ionization potential is sensitive to the UV field 
(see Kozma \& Fransson 1998a). During scattering
the UV photons are shifted toward longer wavelengths, both due to
a pure Doppler shift, but also because of the increased
probability of splitting the UV photons into several photons of
longer wavelengths.  A more accurate treatment of the line
scattering is therefore expected to decrease the importance of
photoionization. 

To probe the effects of photoionization we have therefore
made two model calculations, one without and the other including photoionization.
These models to some extent represent the two extremes. The model 
including photoionization completely ignores the effects of UV scattering, 
while the model without photoionization assumes that all UV photons are 
redistributed to longer wavelengths.
In the model including photoionization, we find that the ejecta are 
photoionized mainly by recombination emission.
At 300 days the dominant contribution to the
photoionization of Fe~II is due to recombination to Fe~III.
The ionization structure of these models is displayed in 
Fig.~\ref{feion}, 
for
the models both with and without photoionization included. This figure shows only
the ionization for iron, which dominates the emission
from SNe~Ia at these late phases. The final output of the calculations 
is the 
detailed evolution of the emission of the ejecta at all wavelengths.
Fig.~\ref{spec} shows the modeled and observed spectra of SN 2000cx at $\sim 360$ 
days past maximum. Convolving the calculated spectra with the filter functions
we also obtain passband light curves from the model.
In Fig.~\ref{phot} we show the light curves from this model, together with our late-time
observations of SN 2000cx.

\section{Discussion}

Although most SNe~Ia are rather homogeneous in their properties,
some are quite different. 
SN~2000cx was clearly photometrically peculiar at early phases (Li et
al. 2001).  
It 
provided a
bad fit to standard supernova light curves around maximum light, and 
showed anomalously blue colors after maximum.  It is not at all clear,
however, how the early appearance correlates with  the later
phases. It has recently been suggested that the peculiarities at
early phases were related to a fast-moving clump just outside the
supernova  photosphere (Branch et al. 2004). This is unlikely to
affect the emission at  later phases. Here we use our late-time 
observations of SN 2000cx for a general discussion of the late-time 
emission from SNe~Ia.  We also note that our modeling
is totally generic, and in no way adapted to fit this particular
supernova. Instead of adjusting input parameters to improve the fits
to the data, we will use the models for a general discussion.  In
Fig.~\ref{92a} we also compare the $V$-band light curve of SN 2000cx with the
light curve of another well-observed SN~Ia, SN 1992A, 
and  show that the late-time behavior of SN 2000cx is not very 
different. A future larger sample of SN~Ia late-time light curves, 
in particular in the near-IR, is
nevertheless needed to assess the applicability of SN 2000cx for a general 
discussion.

\subsection{The importance of near-IR light curves}

The main result of these observations is the constant emission in the
near-IR during the observed epochs (Fig.~6).  
We are not aware of other such systematic
observation of a late-time near-IR light curve of a SN~Ia. In
fact, very little has been published in this respect 
since the sparse, but pioneering, observations
presented by Elias \& Frogel (1983).
Recently, two late-time $H$-band data points for SN~1998bu 
(250--350 days past maximum) suggest that this SN~Ia also
had a flat near-IR light curve at late phases (Spyromilio et al. 2004).

Our constant $J$-band and $H$-band light curves immediately highlight the main
point of our study,  the increasing importance of the near-IR at late phases.  
From our constructed UVOIR light curve, we can investigate where most of the
energy emerges. In Fig.~\ref{fraction} we show the fraction of energy in the
near-IR bands compared to the UVOIR luminosity as it evolves with time.
 This color evolution is also seen in Fig.~7 where the UVOIR
light curve is plotted together with the (arbitrarily shifted) $V$-band light 
curve.  While the $V$ band follows the UVOIR light curve reasonably well
at early phases, at the later epochs the $V$ band declines faster
than the UVOIR curve.

The increasing
importance of the near-IR can also be seen in our detailed
modeling.  In Fig.~\ref{phot} the light curves for the $B$ band through
$H$-band are shown for both observations and models.  We find
generally good agreement between observations and models, with a steeper
slope in the $B$, $V$, $R$, and  $I$ bands, and almost constant
light curves in the $J$ and $H$ bands.
The almost constant emission in the near-IR bands is due to
a shift of emission from iron lines in the optical to the strong near-IR 
[Fe~II] lines, which
dominate the $J$ and $H$ bands.
In this way, the temperature decrease in the ejecta gives a color 
evolution momentarily compensating the decreasing radioactive input in the 
near-IR bands. 

While our observations only cover the UVOIR range, the 
models naturally incorporate the true bolometric light curves. 
Figure~14 shows
a comparison of this bolometric light curve and the $V$-band light curve.
Here we can clearly see that the $V$ band does not follow the bolometric light 
curve at the epochs of our observations. 
For the model without photoionization the
$V$ band drops rapidly around 400--500 days, while for
the photoionization model the drop is significantly less rapid. 
The evolution in the $V$ band reflects the temperature and ionization 
evolution of
the ejecta. The temperatures in the model with photoionization
are higher, and therefore a larger fraction of the luminosity is
emerging in the $V$ band. For the model without photoionization the
emission is emerging at longer
wavelengths.  The lower panel of Fig.~\ref{bolv} 
shows the observed UVOIR light curve
compared with the observed $V$-band light curve (i.e., this is basically a 
close-up of the late phases shown in Fig.~7).
The implications are further discussed below.

\subsection{The Nickel mass}

The simple toy model adopted for the light curve in Fig.~\ref{toymodel} 
provides a reasonable fit to the UVOIR light curve. This exercise reveals 
several potentially interesting aspects of the light curve.

We note that the slope of the four late epochs in the UVOIR light curve is
$\sim1.0$ mag per 100 days. Intriguingly, this is the same as the decay time scale 
of  $^{56}$Co and in the simple model the good match is solely due to 
powering of the 
kinetic energy of completely trapped positrons (dashed curve, Fig.~\ref{toymodel}).
If this interpretation is valid, and if the UVOIR light curve describes the 
true bolometric luminosity of the supernova at these epochs, 
the nickel mass can be directly deduced. The
toy model light curve in Fig.~\ref{toymodel} is from

$L =  1.3 \times 10^{43}\, M_{\rm Ni}~ 
\mathrm{e}^{-t/111.3}  \left(1 - 0.966\mathrm{e}^{-\tau}\right)$,
in units of erg s$^{-1}$, 
where M$_{\rm Ni}$ is the amount of $^{56}$Ni in solar masses and the other
parameters are as defined in section 4 above.
For this simple scenario, only including $^{56}$Co, the parameter 
$t_1$ is related to the time when positrons start to dominate
($t_{pos}$) by $t_{pos}$ = 5.27 $t_1$.
In a least-squares sense, the best fit is for a nickel mass of 0.28~\msun 
and $t_1 = 31.5$ days (fitting all data later than day 50).

Two ways to derive nickel masses from the photometry of SNe~Ia
have been employed (see, e.g., Leibundgut \& Suntzeff 2003
for a review).  Cappellaro et al. (1997) adopted a simple light-curve
model for the late  epochs of the supernova light curve and used the 
$V$ band as a surrogate for the bolometric luminosity.  This is in a way
similar to the discussion above, but we have already noted that at
these late  phases it is not correct to assume the $V$ band to closely follow
the bolometric light curve.

The standard way to derive a nickel mass has rather been to use the
peak luminosity. According to Arnett's rule  (Arnett 1982; Pinto \&
Eastman 2000), the luminosity at the peak directly reflects the  input
powering from the radioactivity. This has been used by Contardo
et al. (2000) and others to derive nickel masses for a number of 
SNe~Ia. The main uncertainty is usually the poorly known
distances and extinctions to the supernovae.

Using the formalism of Contardo et al., the peak UVOIR luminosity in
Fig.~\ref{toymodel} 
corresponds  to $\sim0.46$~\msun~of $^{56}$Ni. In fact, since we
have omitted any flux outside the  0.36--2.2~$\mu$m range, the actual
peak luminosity must be somewhat higher (Candia et al. 2003; Suntzeff
1996).\footnote{The peak luminosity of SN 2000cx from Candia et al. (2003),
log (L/erg~s$^{-1}$) $\approx$ 43.05, corresponds according to Contardo 
et al. (2000) to $\sim$0.56~\msun, but
Contardo et al. also did not include the UV part of the spectrum.
The actual nickel mass of course depends crucially on the assumed distance, 
which is rather uncertain; we adopt 31.2 Mpc.}  

We derive a smaller nickel mass from the late phases compared to
the peak, given the same distance modulus and extinction.
This is likely due to the fact that at these phases, even the UVOIR light curve
does not encapsulate the total bolometric luminosity.
If so, this can then be used to derive the fraction of emission outside the
covered bands.
We thus see that 
our observations indicate that at 500 days about
40\% of the emission is emerging outside our observed range, 
most likely redward of the $K$ band.

This agrees reasonably well with our detailed models, where 
we can see that only about 75\% of the emission at 
400 days is in the UVOIR wavelength range, while the 
mid-IR and far-IR include most 
of the remaining power. Adjusting the derived nickel mass for this recovers
$\sim$0.4~\msun~of $^{56}$Ni.

The late-time UVOIR slope points to complete (or at least constant) 
trapping of positrons,
and  gives no hint of temporal color evolution out of the observed
bands. The models do in fact show a slow evolution, in that the UVOIR range
progressively lose energy to the far-IR, but this is also one of the aspects 
where the model is clearly not completely correct (see Sect. 6.3).

In our models we have used a nickel mass of 0.6~\msun~from the W7 model, 
and the same distance modulus and reddening as adopted for SN 2000cx. 
However, the fits to the data in Fig.~\ref{phot} are not really good enough to determine
the nickel mass with any precision. A lower nickel mass would clearly
enhance the rapid luminosity drop, but this is also affected by the density
(i.e., the assumed ejecta mass and expansion velocity).

\subsection{The IR catastrophe}

The significance of a thermal instability for SNe~Ia at
late phases was recognized by Axelrod (1980). At low temperatures the
cooling of the ejecta is rapidly shifted from the optical into the IR
fine-structure lines (e.g., [Fe~I] 24~$\mu$m, [Fe~II] 26~$\mu$m).  This
so-called ``infrared catastrophe'' would drastically decrease the flux in the optical.
Fransson et al. (1996) 
also found a rapid decrease in the
optical flux at $\sim500$ days past explosion in their modeling, and
realized that this was not in accordance with late-time optical
observations of SN 1972E.  Now, 10 years later, the accuracy of the
atomic data needed for these models, as well as the computing
facilities, have improved significantly. 
Still, our new models show the
same effect. 

In Fig.~\ref{phot} it can be seen that
the curves for the model without 
photoionization 
drops quickly after 
400--600 days. 
The faster drop in this model is due to a lower temperature 
in the ejecta.
Again, this rapid drop in flux is in conflict with the
late-time observations (in particular the very last epoch).  
Since these models attempt to include our best
knowledge of the physics of the emission of the supernova, the reason
for this discrepancy is important to understand. 

When looking at the ionization structure for the two models at 300 days
(Fig.~\ref{feion}), 
we find that the degree of ionization is significantly higher in the
photoionization model. Especially in the outermost region (where the
iron abundance is small) the amount of Fe III, at the expense of Fe II,
increases when the photoionization is turned on.
It is the
recombination emission (to Fe III) from the underlying regions
that photoionizes Fe II. In both models the amount of Fe I is negligible
at this epoch.
In fact, although our spectral calculations based on these 
ionization structures give a decent 
fit to the observed spectrum
(Fig.~\ref{spec}), 
it appears as if the model with lower ionization does the best job.
The light curve models in Fig.~\ref{phot} 
do instead favor the models including 
photoionization.
A proper treatment of the UV scattering is thus clearly important, but 
since the differences in the light-curve slopes are mainly
a temperature effect there are other 
scenarios which might provide similar results.
To avoid the IR catastrophe there is basically a need to 
keep the ejecta warmer than a critical 
temperature where this instability sets in. This can be obtained in several 
ways.
For example, a higher nickel mass would give higher temperature and 
ionization. Another possible mechanism is 
clumping of the ejecta, which would give regions of lower and
higher densities, and the low-density regions would be hotter due to less 
efficient cooling. Yet another scenario which would affect
the temperature structure in the ejecta is the non-local deposition
of positrons.  These matters are beyond the scope of the current 
investigation and will be discussed elsewhere.

\subsection{Positron escape}

The above results demonstrate the difficulty in interpreting supernova light 
curves for which observations only exist over a limited wavelength range.

Milne et al. (2001) used a Monte Carlo scheme to simulate the deposition of 
$\gamma$-rays and positrons in the ejecta. They claim that late light curves of 
SNe Ia can be reproduced only if a substantial fraction of the positrons 
escape from the ejecta. 
They based that claim upon the suggestion that at 
late phases ($\gtrsim$ 40 days), the $V$ band constitutes 
a constant ($\sim$25\%) fraction of the 
3500--9700~\AA\ emission, which they loosely referred to as the bolometric 
luminosity. We show that by including UVOIR photometry 
in the determination of a bolometric luminosity, at late 
phases the $V$ band is seen to deviate from the bolometric luminosity 
for SN 2000cx. 

In Fig. \ref{bolv} we compare the bolometric light curve to the $V$-band 
light curve for our
model calculations and observations. 
These model calculations assume
instant and local deposition of the positrons. 
Even in this case of full positron trapping we find an increasing 
deviation between the bolometric and $V$-band light curve
with time. 
The lower panel of Fig. \ref{bolv} shows our observations, where 
we also 
find that the slope of the bolometric and $V$-band
light curves differs. A comparison to a true bolometric light curve 
would further increase this effect.

It is clear that this shift of the emission into the IR passbands
with time will mimic the effect of positron escape. Deviations between 
simple models and the $V$-band light curve 
should therefore not be 
interpreted in terms of the degree of positron escape. Instead, a
detailed and
consistent knowledge of the temperature and ionization evolution of
the ejecta is required.  This  means that many previous studies of positron
escape from the late-time ejecta in  SNe~Ia in fact require
more detailed observations and modeling than hitherto appreciated
to draw definitive conclusions on positron trapping.

\subsection{The very late-time light curve}

Lastly, we note that the $V$-band decline appears to level out in the
two  final {\it HST} observations; the slope is only 0.65 mag per 100 days
between day 557 and 693 in the F555W {\it HST} observations.  It is
hazardous to draw any general conclusions from a single passband since there
are many ways to power the late-time light curve of a supernova (see, e.g.,
Sollerman et al. 2002). This section is therefore rather speculative.

First of all, there could of course be a late-time light echo, as seen for 
SN~1998bu. However, the SO host galaxy should be clear of dust and the {\it HST}
images show no evidence for a ring. We note also that the SN~Ia 
1992A displayed a slow F555W evolution at these late phases 
(Cappellaro et al. 1997). 
This behavior may thus be generic for thermonuclear supernovae.

The final bright $V$-band observation immediately shows that the
IR catastrophe as indicated in the models does not occur at
these phases. As noted previously, there are many possible solutions
to this dilemma.

Potentially, $^{44}$Ti could become important at 
late phases, since it
decays on a time scale of 87 years, 
and 
the decay
mainly results in positrons. 
However, to influence the light curve at our last epoch of observations 
requires a large abundance of titanium --- as much as 7\% of the original 
mass of $^{56}$Ni. 
This is much more than in the W7 model, which we have used for 
our detailed calculations, and is likely ruled out from nucleosynthesis 
arguments at the temperatures reached in SN Ia burning fronts.

We emphasize that this late observation is only in the $V$ band, and
is not representing the bolometric luminosity. Thus, any effect that
could boost the  emission in this band relative to the other bands
would give a shallower slope in the $V$ band. 
In our model calculations we find that after $\sim650/800$ days 
(with/without photoionization)  Fe~I dominates the emission in the $V$ band.
Fe~I emits relatively more in the $V$ band and this results in a flattening of 
the $V$-band light curve (see also Fransson et~al. 1996).
However, even though our modeled slope of the $V$-band light curve levels out, 
the total flux of Fe~I is clearly too low compared to the observations. 
Also, the epoch at which Fe~I starts to dominate is too late. 
A model with a clumpy medium could possibly overcome these problems.
Another possible effect is indeed 
a longer lifetime of the positrons that allows time-dependent effects 
to become dominant.
This is very speculative and more very late-time
observations of SNe~Ia are needed before any serious
discussion can be made.

\section{Summary}

We have conducted a systematic and comprehensive monitoring programme
of the SN~Ia 2000cx at late phases using the VLT and
{\it HST}. The VLT observations cover phases 360 to 480 days past maximum
and includes photometry in the $(U)BVRIJH(K)$ bands.  While the optical
bands decay by about 1.4 mag per 100 days, we find that the
near-IR magnitudes stay virtually constant during the observed
period. This means that the importance of the near-IR to the
bolometric light curve increases with time. The finding is also in
agreement with our detailed modeling of a SN~Ia in the
nebular phase. In these models, the increased contribution of the
near-IR is simply a temperature effect.  We note that this complicates
late-time studies where often only the $V$ band is well monitored. In
particular, it is not correct to assume that any optical band follows
the bolometric light curve at these phases, and any conclusions based
on such assumptions must be regarded as premature.

We have found that a very simple model where all positrons are trapped
can reasonably well account for the UVOIR observations. The nickel
mass deduced from the positron tail of the light curve is
lower than found from the  peak brightness, which provides the fraction of 
emission outside the observed range at the late phases.
Our detailed models show the signature of an infrared catastrophe at these 
epochs,
which is not supported by the observations. More importantly, at the
observed epochs, the gradual lowering in temperature increases the
importance of the near-infrared emission.  These conclusions are drawn from
observations of a single supernova, which was clearly unusual at the
peak. We support this generalization by detailed modeling and
comparisons to the late-time light curve of SN 1992A. It is clear, however,
that these suggestions have to be verified by more data of
thermonuclear supernovae at very late phases. Such a programme is
currently ongoing.  The investigation of positron escape from SNe~Ia
thus awaits observations that better represent the bolometric
luminosity as well as more realistic  models.\\

\begin{acknowledgements}
We are grateful to Ken Nomoto for providing the W7 model, and to Sultana Nahar
for providing recombination rates for the iron ions.  
We also thank Nick Suntzeff for important input in the early phases of 
this project.
We thank the ESO VLT
staff for their assistance with the observations.  
A.V.F. is grateful for NSF grants AST-9987438 and
AST-0307894, as well as for NASA grants GO-8602 and GO-9114 from the Space
Telescope Science Institute (STScI), which is operated by AURA, Inc., under
NASA contract NAS 5-26555.

\end{acknowledgements}

\clearpage

\begin{table}[b]
\caption{Log of optical VLT observations, SN~2000cx.}
\label{t:obs}
\begin{tabular}{llllll}
\hline
\hline
Date & Filter  & MJD$^a$ & Exposure & Airmass$^a$ & Seeing  \\ 
(UT) &         &         &   (s)    &            & (arcsec) \\
\hline

2001 07 21  & $B$ & 52111.36 & 2x300 & 1.31 & 1.05 \\
2001 07 21  & $V$ & 52111.37 & 2x240 & 1.29 & 0.98 \\
2001 07 21  & $R$ & 52111.38  & 2x240 & 1.27 & 0.91 \\
2001 07 21  & $I$ &  52111.38 & 2x500 & 1.25 & 0.90 \\
2001 07 21  & $U$ & 52111.40  & 2x780 & 1.23 & 1.05 \\
2001 07 22  & $V$ & 52112.34  & 2x300 & 1.38 & 0.91 \\

2001 09 13 & $V$ & 52165.25  & 3x300 & 1.23 & 0.94 \\
2001 09 13 & $I$ & 52165.26  & 3x800 & 1.22 & 0.86 \\
2001 09 13 & $B$ & 52165.29 & 3x400 & 1.21 & 0.92 \\
2001 09 13 & $R$ & 52165.31 & 4x400 & 1.23 & 0.87 \\

2001 10 10 & $R$ & 52192.13 & 750+418 & 1.34 & 0.93 \\

2001 10 17 & $I$ & 52199.15 & 8x650 & 1.24 & 0.68 \\
2001 10 17 & $B$ & 52199.22 & 4x750 & 1.24 & 0.67 \\
2001 10 17 & $V$ & 52199.25 & 4x750 & 1.35 & 0.86 \\
2001 10 19 & $R$ & 52201.22 & 4x750 & 1.25 & 0.86 \\
2001 11 19 & $I$ & 52232.02 & 8x650 & 1.38 & 0.92 \\
2001 11 19 & $R$ & 52232.08 & 4x750 & 1.21 & 0.78 \\
2001 11 20 & $V$ & 52233.02 & 4x750 & 1.36 & 0.84 \\
2001 11 20 & $B$ & 52233.06 & 4x750 & 1.24 & 0.79 \\

\hline
\end{tabular} \\
\begin{tabular}{lll}
$^a$ Refers to the first exposure. && \\
\end{tabular}
\end{table}

\begin{table}[b]
\caption{Magnitudes for local standards in the optical.}
\label{locals}
\begin{tabular}{lllllll}
\hline
\hline
Offsets$^a$& & $U$ & $B$ & $V$ & $R$ & $I$ \\
    &  &   &   &   &   &  \\
\hline
115.12~W& 17.86~N & 23.69 (0.18)$^b$  & 23.29 (0.04) &  22.45 (0.02)  &  22.00 (0.01) &  21.48 (0.02) \\
134.48~E& 154.34~S & 21.59 (0.03)  & 21.54 (0.03) &  20.77 (0.02)  &  20.29 (0.01) &  19.76 (0.01) \\
140.48~E& 88.12~S & 24.16 (0.20)  & 23.04 (0.03) &  21.48 (0.02)  &  20.58 (0.01) &  19.55 (0.01) \\
52.08~W& 116.12~S & 22.59 (0.06)  & 21.94 (0.02) &  20.86 (0.02)  &  20.23 (0.01) &  19.57 (0.01) \\
40.55~W& 106.03~S & 22.73 (0.07)  & 22.62 (0.04) &  21.80 (0.01)  &  21.30 (0.01) &  20.73 (0.02) \\
158.45~E& 45.25~N & 22.40 (0.07)  & 23.17 (0.04) &  22.38 (0.01)  &  21.84 (0.02) &  21.29 (0.03) \\
156.22~E& 187.75~N & 21.52 (0.03)  & 21.68 (0.01) &  21.18 (0.02)  &  20.85 (0.02) &  20.47 (0.01) \\
188.73~E& 6.06~S  & 24.13 (0.22)  & 23.04 (0.02) &  21.60 (0.02)  &  20.72 (0.02) &  19.67 (0.01) \\
24.84~W& 12.28~N  & 24.51 (0.29)  & 23.59 (0.03) &  22.69 (0.03)  &  22.15 (0.02) &  21.49 (0.02) \\
62.29~W& 9.51~N   & 24.55 (0.28)  & 23.88 (0.04) &  22.97 (0.03)  &  22.45 (0.02) &  21.83 (0.03) \\
68.48~W& 121.36~S & 23.01 (0.10)  & 23.09 (0.02) &  22.15 (0.02)  &  21.60 (0.01) &  20.95 (0.01) \\
109.51~W& 180.10~N & 23.77 (0.15)  & 23.57 (0.03) &  22.86 (0.03)  &  22.39 (0.02) &  21.86 (0.02) \\
88.22~W& 78.11~S & 23.02 (0.09)  & 23.08 (0.02) &  22.34 (0.02)  &  21.88 (0.02) &  21.32 (0.02) \\
82.48~E& 104.51~S & 23.86 (0.18)  & 24.05 (0.04) &  22.92 (0.03)  &  22.01 (0.02) &  21.18 (0.02) \\
\hline
\end{tabular} \\
\begin{tabular}{lll}
$^a$ \ Offsets in arcseconds measured from the supernova.\\
$^b$ \ Numbers in parentheses are uncertainties && \\
\end{tabular}
\end{table}

\begin{table}[b]
\caption{Late-time optical magnitudes of SN~2000cx.}
\label{optmags}
\begin{tabular}{lllllll}
\hline
\hline
MJD$^a$ & Phase$^b$ & $U$ & $B$ & $V$ & $R$ & $I$ \\ 
52000+  &  (days)  &   &  & & &           \\
\hline

111.4       & 359 & 22.30 (0.08) & 21.27 (0.03) & 21.28 (0.01) & 22.17 (0.02) & 21.70 (0.03) \\
112.3       & 360 &              &              & 21.28 (0.03) &              &              \\
165.3       & 413 &              & 22.00 (0.03) & 22.02 (0.03) & 22.91 (0.03) & 22.09 (0.03) \\
192.1       & 440 &              &              &              & 23.26 (0.04) &              \\
199.2       & 447 &              & 22.50 (0.03) & 22.46 (0.01) &              & 22.42 (0.03) \\
201.2       & 449 &              &              &              & 23.46 (0.03) &              \\
232.5       & 480 &              & 23.00 (0.04) & 23.03 (0.02) & 23.85 (0.06) & 22.84 (0.05) \\

\hline
\end{tabular} \\
\begin{tabular}{lll}
$^a$ Refers to the first exposure. && \\
$^b$ Days past maximum B-band light. && \\
\end{tabular}
\end{table}

\clearpage

\begin{table}[b]
\caption{Log of near-IR observations, SN~2000cx.}
\label{irmags}
\begin{tabular}{llllll}
\hline
\hline
Date & Filter  & MJD$^a$ & Exposure$^b$ & Airmass$^a$ & Seeing  \\ 
(UT)   &         &         &   (s)      &            & (arcsec) \\
\hline

2001 07 31 & $J$  & 52121.35 & 30x3x13 & 1.25 & 0.61 \\
2001 07 31 & $H$  & 52121.37 & 15x6x13 & 1.22 & 0.40 \\
2001 07 31 & \ks\  & 52121.40 & 12x6x16 & 1.21 & 0.39 \\

2001 09 21 & $J$  & 52173.22 & 30x3x13 & 1.23 & 0.49 \\
2001 09 21 & $J$  &  & 30x3x15 &  & 0.49 \\
2001 09 21 & $H$  & 52173.26 & 15x6x13 & 1.21 & 0.40 \\
2001 09 21 & $H$  &  & 15x6x14 &  & 0.40 \\

2001 11 19 & $J$  & 52232.01 & 30x3x15 & 1.40 & 0.63 \\
2001 11 19 & $J$  &  & 30x3x13 &  & 0.63 \\
2001 11 19 & $H$  & 52232.05 & 15x6x13 & 1.25 & 0.48 \\
2001 11 19 & $H$  &  & 15x6x14 &  & 0.48 \\

2001 11 20 & $J$  & 52233.01 & 30x3x15 & 1.42 & 0.60 \\
2001 11 20 & $J$  &  & 30x3x13 &  & 0.60 \\
2001 11 20 & $H$  & 52233.05 & 15x6x13 & 1.26 & 0.44 \\
2001 11 20 & $H$  &  & 15x6x14 &  & 0.44 \\

2001 11 22 & $J$  & 52235.01 & 30x3x15 &  1.38 & 1.00 \\
2001 11 22 & $J$  &  & 30x3x13 &  & 1.00 \\
2001 11 22 & $H$  & 52235.05 & 15x6x13 & 1.24 & 0.68 \\
2001 11 22 & $H$  &  & 15x6x14 &  & 0.68 \\

\hline
\end{tabular} \\
\begin{tabular}{lll}
$^a$ Refers to the first exposure. && \\
$^b$ DITxNDITxNEXP && \\
\end{tabular}
\end{table}

\begin{table}[b]
\caption{Magnitudes of local standards in the near-IR.}
\label{localsIR}
\begin{tabular}{rllll}
\hline
\hline
Offsets$^a$ & & $J$ & $H$ & \ks\ \\
       & &   &   &    \\
\hline
5.34 E& 2.23 S &  18.10 (0.06)   & 17.47 (0.03)     &  17.15  (0.01)     \\
17.04 W& 18.25 S &  20.23 (0.08)  &  19.22 (0.06)     &  18.35  (0.03)      \\
80.87 W& 17.36 N &  18.45 (0.06)   & 17.84 (0.02)     &  17.57  (0.02)      \\
33.98 W& 26.71 N &  19.87 (0.05)   & 19.07 (0.04)     &  18.77   (0.04)     \\
36.80 W& 21.81 N &  18.30 (0.05)  &  17.66  (0.03)    &  17.29   (0.02)     \\
83.25 W& 70.93 N &  16.08 (0.04)   & 15.42 (0.01)     &  15.14   (0.01)      \\
105.06 W& 40.36 N &  17.54 (0.05)  &  17.12  (0.02)    &  17.02   (0.01)      \\
130.73 W& 29.53 N &  19.03 (0.06)  &  18.31  (0.03)    &  18.10   (0.02)       \\
103.50 W& 56.17 N &  20.37 (0.07)   & 19.72  (0.07)    &  19.55   (0.06)      \\
49.27 W& 12.32 N &  18.29 (0.06)  &  18.14  (0.03)    &  17.87   (0.02)      \\
97.35 W& 66.92 N &  21.13 (0.07)   & 20.47  (0.07)    &  20.07   (0.09)      \\
115.60 W& 17.95 N &  20.94 (0.06)  &  20.46  (0.07)    &  20.23   (0.10)        \\
37.25 W& 25.82 N &  20.37 (0.07)  &  19.70  (0.05)    &  19.50   (0.06)       \\
57.13 W& 40.81 N &  20.01 (0.07)  &  19.04  (0.04)    &  18.14   (0.02)      \\

\hline
\end{tabular} \\
\begin{tabular}{lll}
$^a$ Offsets in arcseconds measured from the supernova. && \\
\end{tabular}
\end{table}

\clearpage

\begin{table}[b]
\caption{Late-time near-IR magnitudes of SN~2000cx.}
\label{IRmags}
\begin{tabular}{lllllll}
\hline
\hline
MJD$^a$ & Phase$^b$ & $J$ & $H$ & \ks\  \\ 
52000+  &  (days)   &   &   &\\
\hline

121.4    & 369 & 21.86 (0.08)$^c$ & 21.33 (0.13) &  20.32 (0.11)&\\
173.3    & 421 & 21.83 (0.07) & 21.05 (0.10) &   &\\
232.0    & 480 & 21.87 (0.09) & 21.15 (0.14) &   &\\
233.0    & 481 & 21.73 (0.09) & 20.95 (0.12) &   &\\
235.0    & 483 & 21.77 (0.11) & 20.99 (0.11) &   &\\

\hline
\end{tabular} \\
\begin{tabular}{lll}
$^a$ Refers to the first exposure. && \\
$^b$ Days past maximum B-band light. && \\
$^c$ Numbers in parentheses are uncertainties. However, as 
discussed in Sect. 2.2 the total IR uncertainties \\
for the faint supernova are likely as large as 0.15 mags at these epochs.  
\end{tabular}
\end{table}

\begin{table}[b]
\caption{{\it HST} observations of SN~2000cx.}
\label{hsttable}
\begin{tabular}{llllll}
\hline
\hline
Date & Phase$^a$ & Filter &  Magnitude & Exposure & PI  \\ 
(UT) &  (Days)   &        &            & (s)        &    \\
\hline
2001 07 10 & 348 & F675W & 21.98$\pm$0.06 & 280 & Filippenko\\
2001 07 10 & 348 & F814W & 21.41$\pm$0.06 &280 & Filippenko\\
2002 02 03 & 557 & F439W  & 24.38$\pm$0.25 & 2100 & Kirshner\\
2002 02 03 & 557 & F555W  & 24.32$\pm$0.08 & 2100 & Kirshner\\
2002 06 19 & 693 & F555W & 25.20$\pm$0.17 & 4200 & Kirshner\\

\hline
\end{tabular} \\
\begin{tabular}{lll}
$^a$ Days past $B$-band maximum light && \\
\end{tabular}
\end{table}

\begin{table}[b]
\caption{Decline rates in the VLT data$^a$}
\label{slopes}
\begin{tabular}{llllll}
\hline
\hline
$B$ & $V$ & $R$ & $I$ & $J$ &  $H$  \\
1.42 (0.04) & 1.38  (0.01) &  1.40  (0.03) &  0.88 (0.04) & $-$0.06 (0.15) & $-$0.23 (0.15) \\
\hline
\end{tabular} \\
\begin{tabular}{lll}
$^a$ mag per 100 days between 360 and 480 days; errors in parenthesis are $1\sigma$. && \\
\end{tabular}
\end{table}

\clearpage

\begin{figure}[t]
\includegraphics[width=80mm,clip]{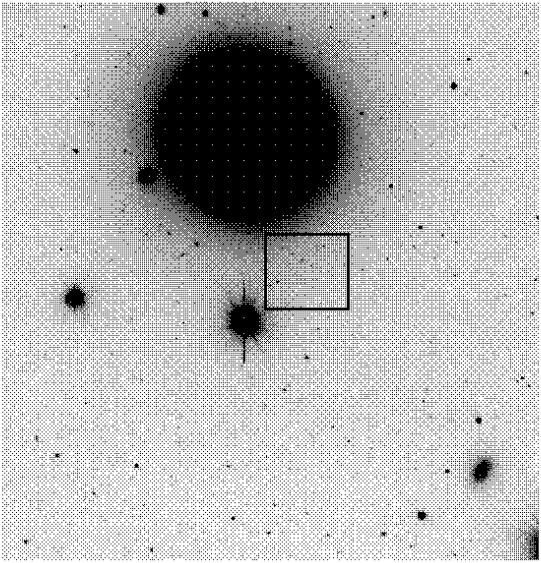}
\caption{$V$-band image obtained with VLT/FORS on 21 July 2001, 359 days past 
maximum light. This is basically the full FORS field of view, 
$6.8' \times 6.8'$. North is up and east to the left. The supernova 
is 23$\arcsec$ west and $109\arcsec$ south of the nucleus of 
NGC~524. The region containing the supernova is marked by a rectangle, 
and shown in detail in Fig.~2.}
\label{FORSimage}
\end{figure}

\clearpage

\begin{figure}[t]
\setlength{\unitlength}{1mm}
\begin{picture}(80,80)(0,0)
\put (  0, 0)    {\includegraphics[width=37mm,bb=142 246 470 547,clip]{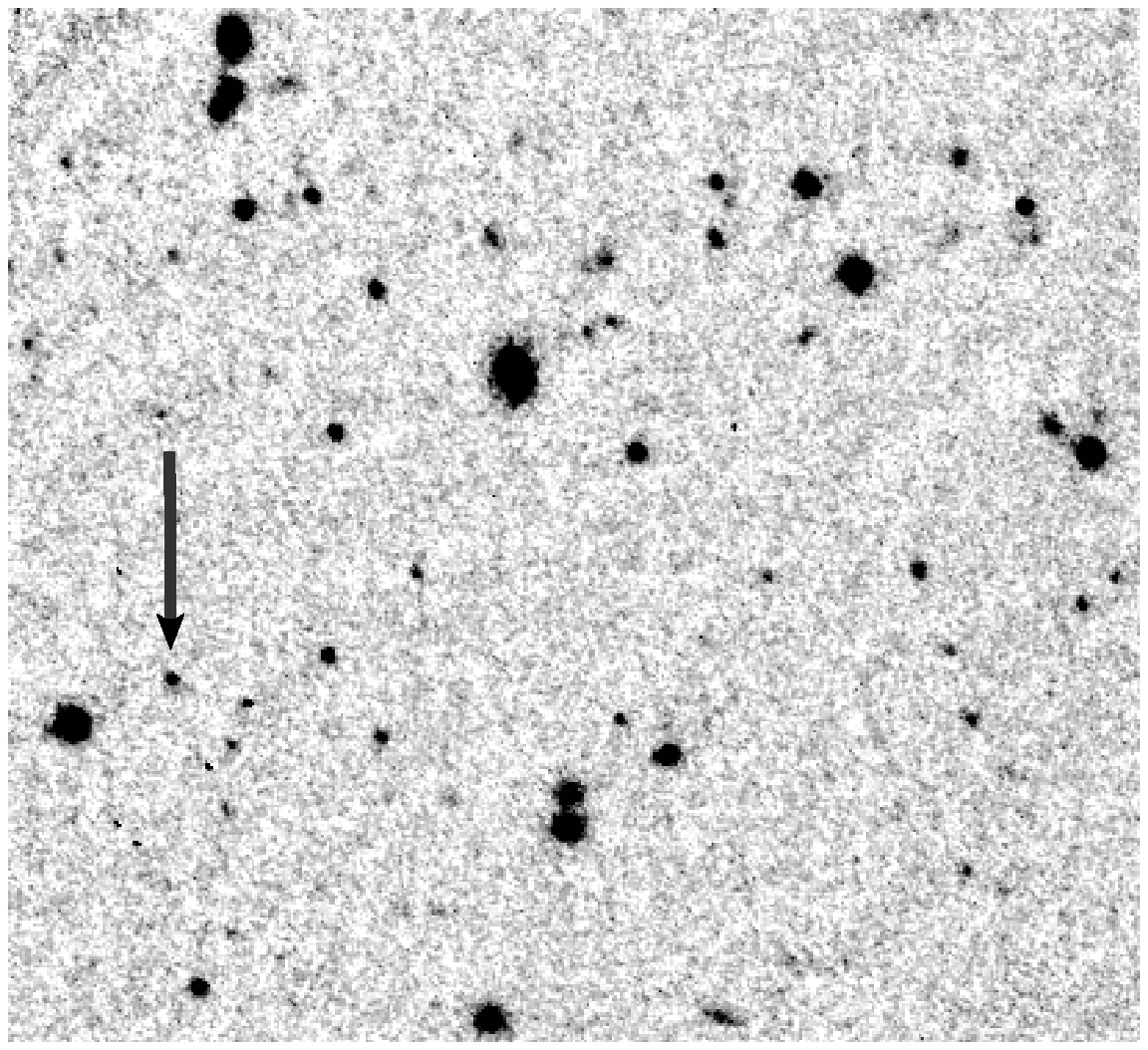}}
\put (  38, 0)   {\includegraphics[width=37mm,bb=142 246 470 547,clip]{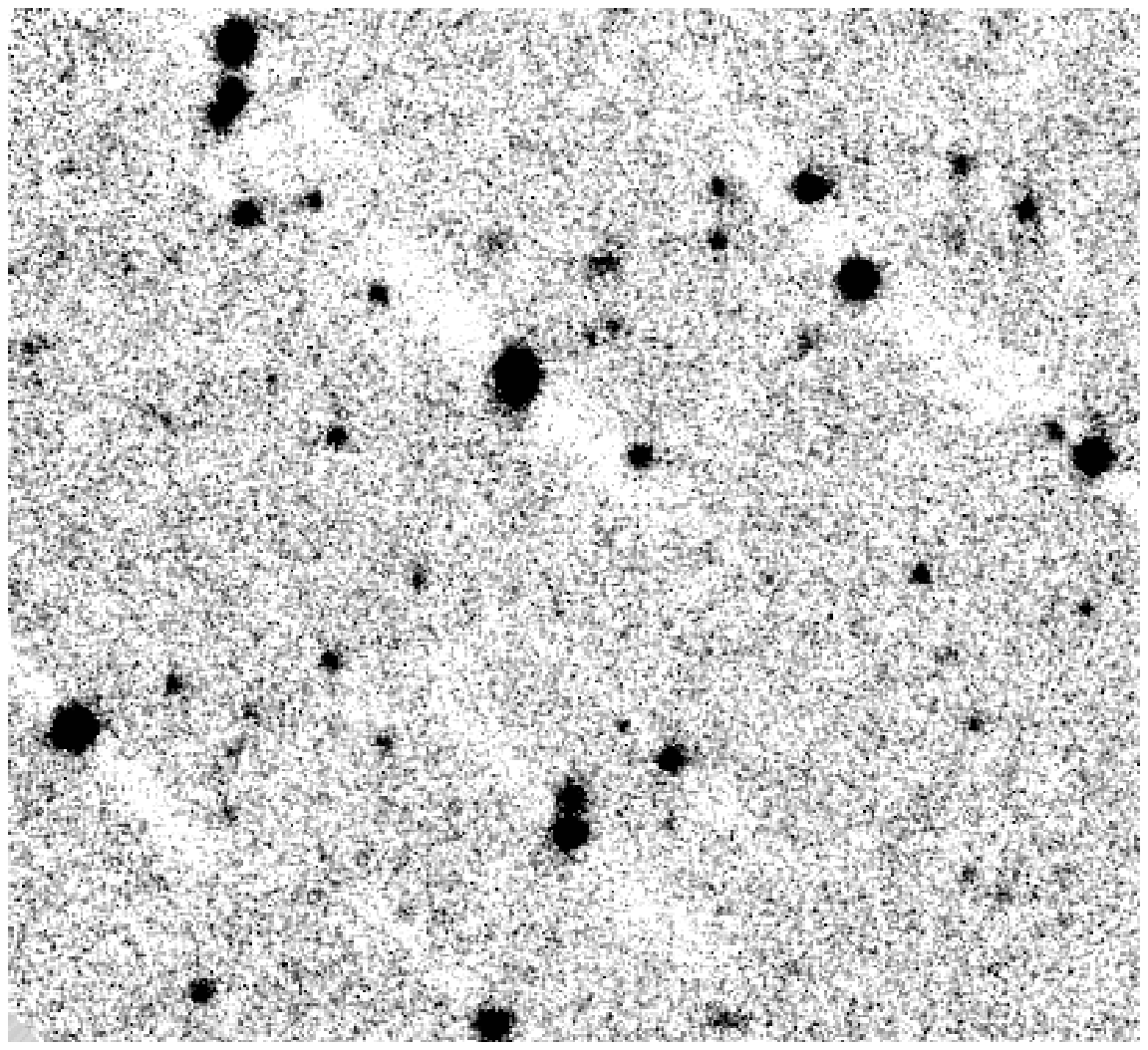}}
\put ( 0,38)     {\includegraphics[width=37mm,bb=63 174 550 619,clip]{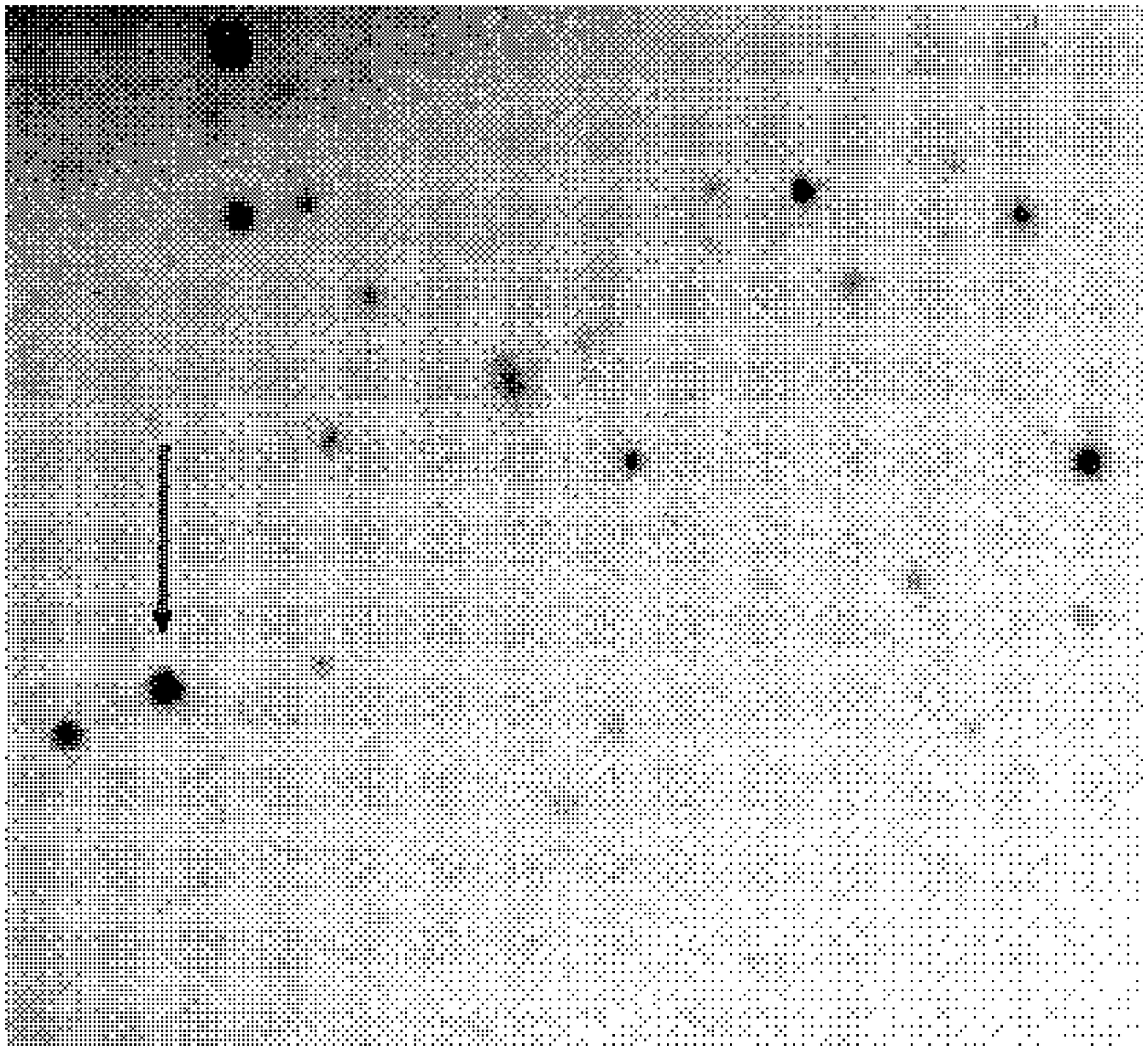}}
\put ( 38, 38)   {\includegraphics[width=37mm,bb=190 290 422 502,clip]{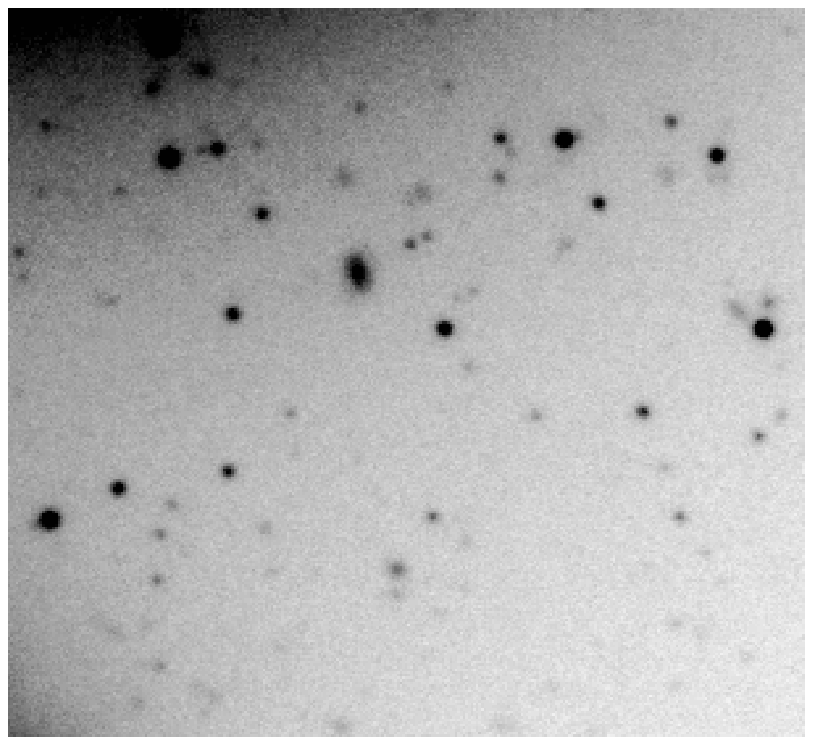}}
\end{picture}
\caption{{\it Upper left:}
Close-up of the region indicated in Fig.~1, $V$ band at 359 days past 
maximum. 
The field of view is $60'' \times 55''$. North is up 
and east to the left. SN 2000cx is marked with an arrow. 
{\it Upper right:} $V$-band image obtained 480 days past maximum. 
Note how the supernova has faded considerably compared to the nearby stars.
{\it Lower left:}
$J$-band image obtained with VLT/ISAAC 369 days past maximum light. 
The orientation and field of view is the same as for the optical image above.
The supernova is marked with an arrow. {\it Lower right:} $J$-band image 
obtained 480 days past maximum. Note that the supernova has not faded since
the previous $J$-band image.
}
\label{blowupimages}
\end{figure}

\clearpage

\begin{figure*}[t]
\includegraphics[width=150mm,clip]{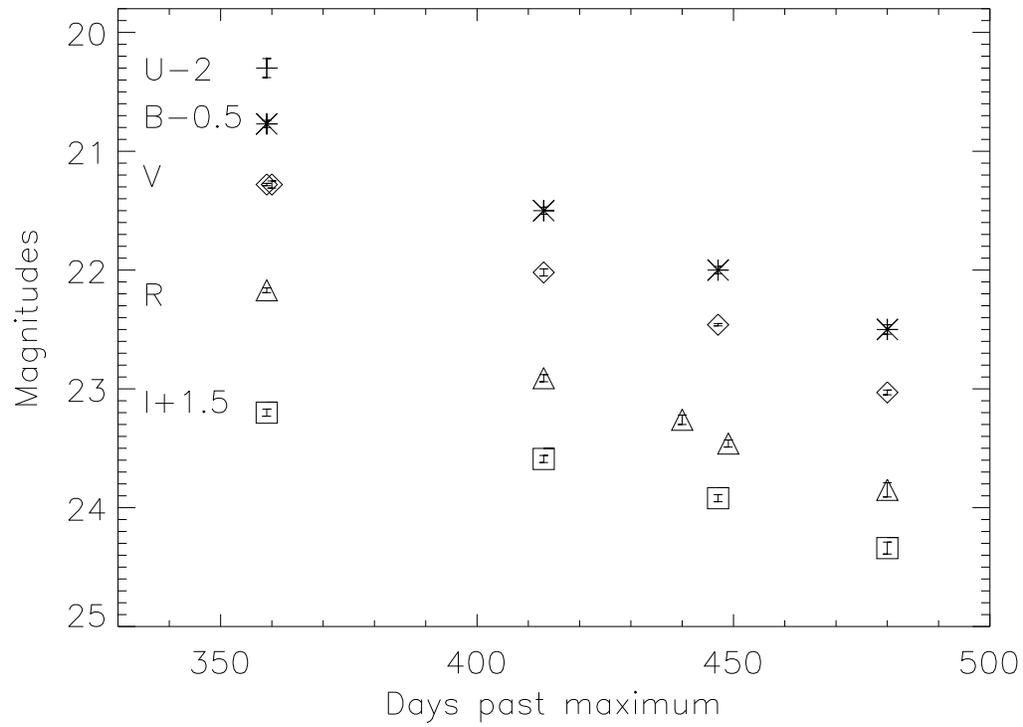}
\caption{Late-time optical light curves of SN 2000cx; see Table 3.
The magnitudes have been shifted for clarity 
as indicated.}
\label{opticallightcurve}
\end{figure*}

\begin{figure*}[t]
\includegraphics[width=150mm,clip]{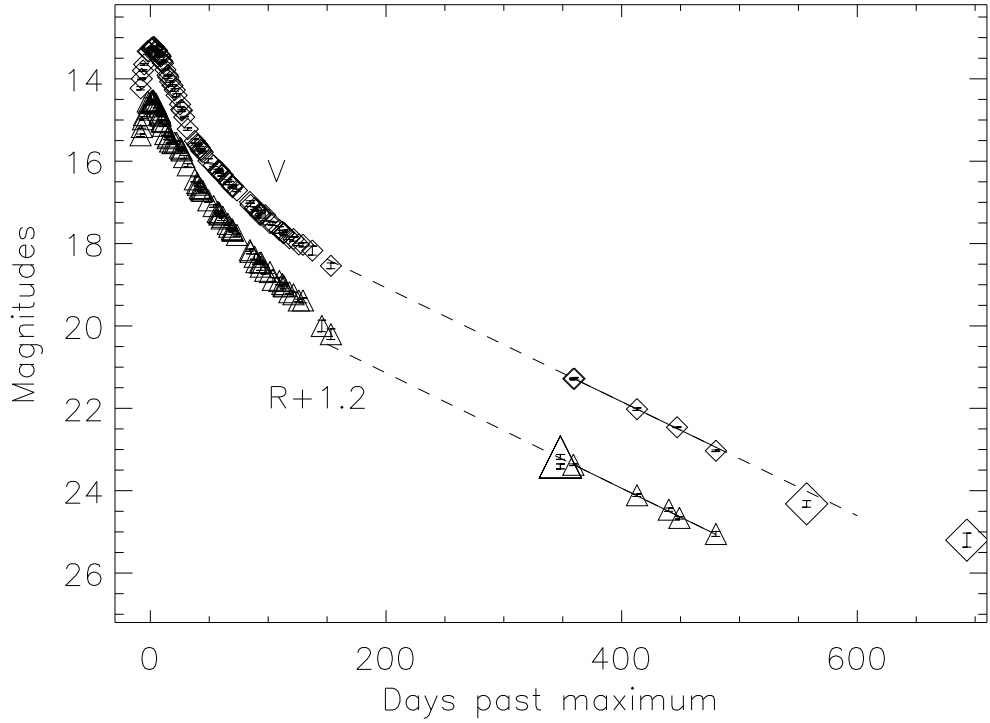}
\caption{The optical light curves of SN 2000cx. The early data are from 
Li et al. (2001). The $R$-band light curve has been shifted as indicated 
for clarity. The full lines are the best linear fits to the late-time VLT 
data, 
with slopes as given in Table~8. The dashed line is an extrapolation of 
this slope back to the early light curve.
The late {\it HST} photometry is displayed with
larger symbols. 
For the $R$-band point at 348 days past maximum, we have indicated 
the likely transformation from {\it HST} magnitudes to the $R$-band system 
(lower error bar).
}
\label{opticallightcurve_li}
\end{figure*}

\begin{figure*}[t]
\includegraphics[width=150mm,clip]{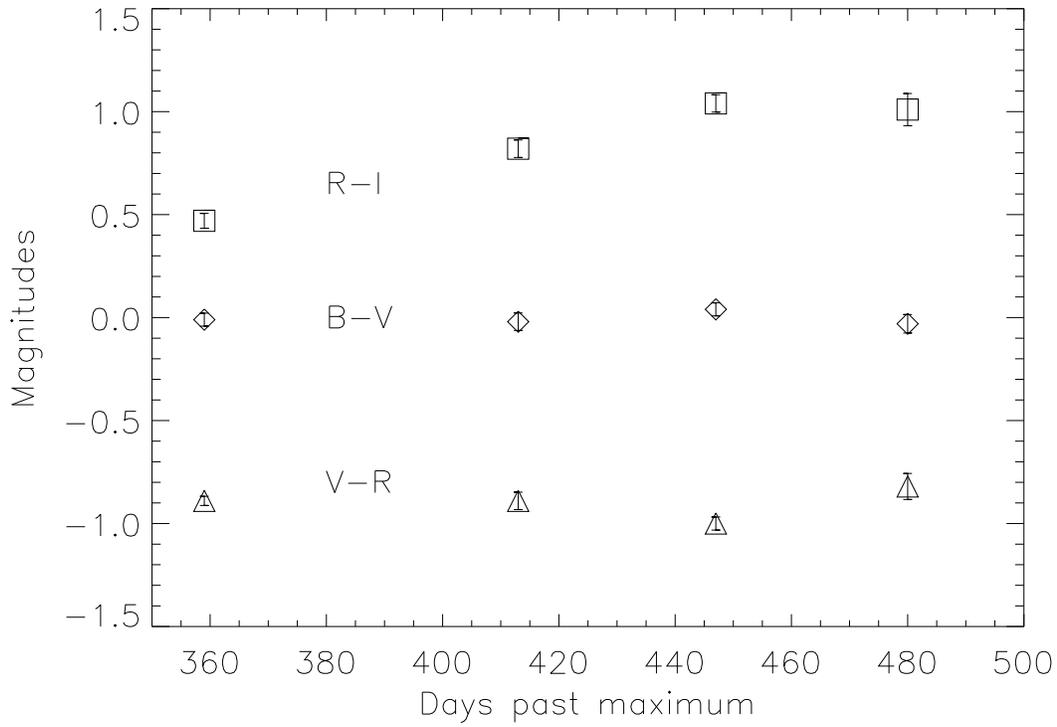}
\caption{Color curves for the late-time optical phases. There is little 
evolution in $B-V$ and $V-R$, while $R-I$ increases with time due 
to the slower decline rate in the $I$ band.}
\label{colors}
\end{figure*}

\begin{figure*}[t]
\includegraphics[width=150mm,clip]{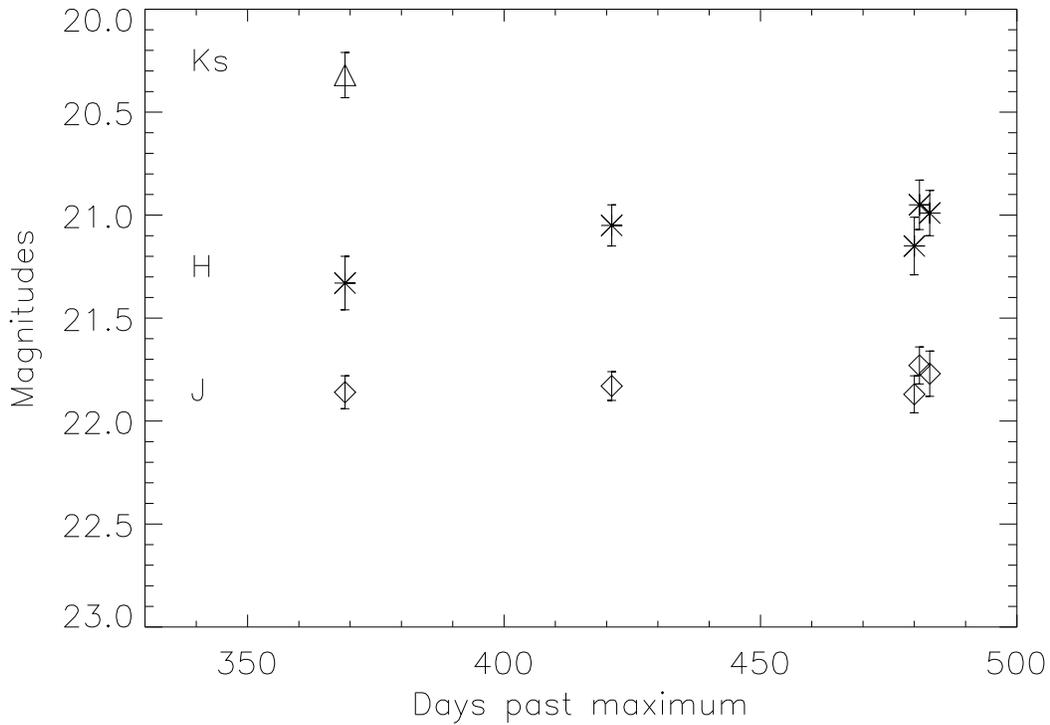}
\caption{Near-IR magnitudes and uncertainties (see Table 6) for SN 2000cx 
at late phases.}
\label{lateIRlc}
\end{figure*}


\begin{figure*}[t]
\includegraphics[width=150mm,clip]{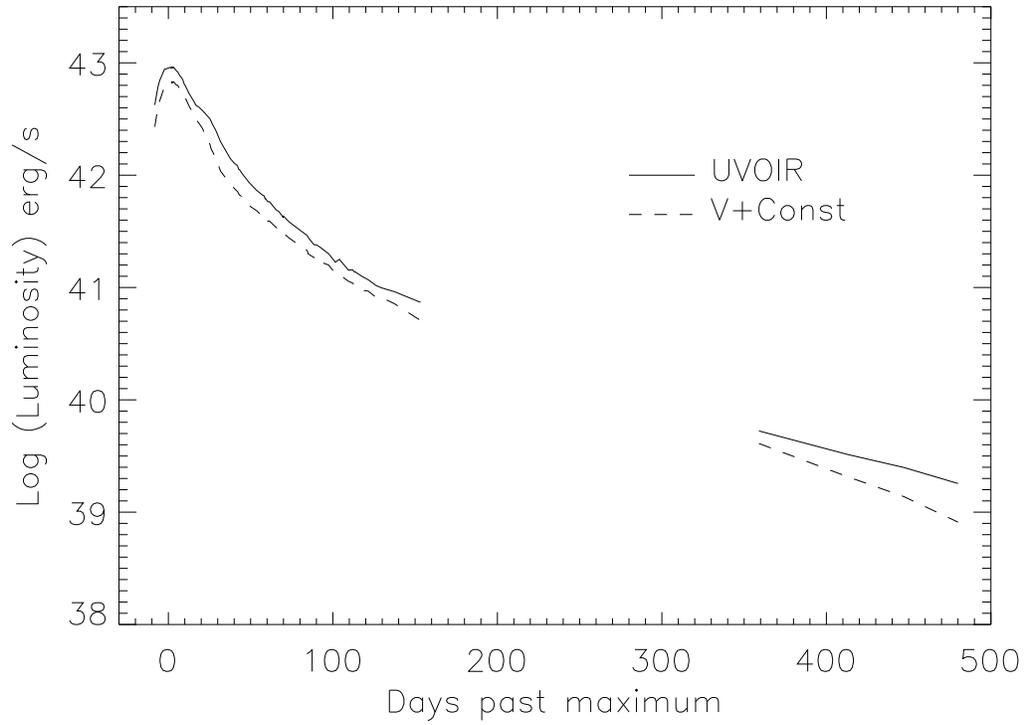}
\caption{UVOIR light curve for SN 2000cx. The early-epoch data are from 
Li et al. (2001, $(U)BVRI$) and Candia et al. (2003, $JHK$). The UVOIR curve 
is shown with the full line, while the dashed line is the $V$-band luminosity
shifted arbitrarily along the ordinate. The details and assumptions for the 
construction of the UVOIR light curve are given in the text. The adopted 
distance is 31.2 Mpc.}
\label{bolometric}
\end{figure*}

\begin{figure*}[t]
\includegraphics[width=150mm,clip]{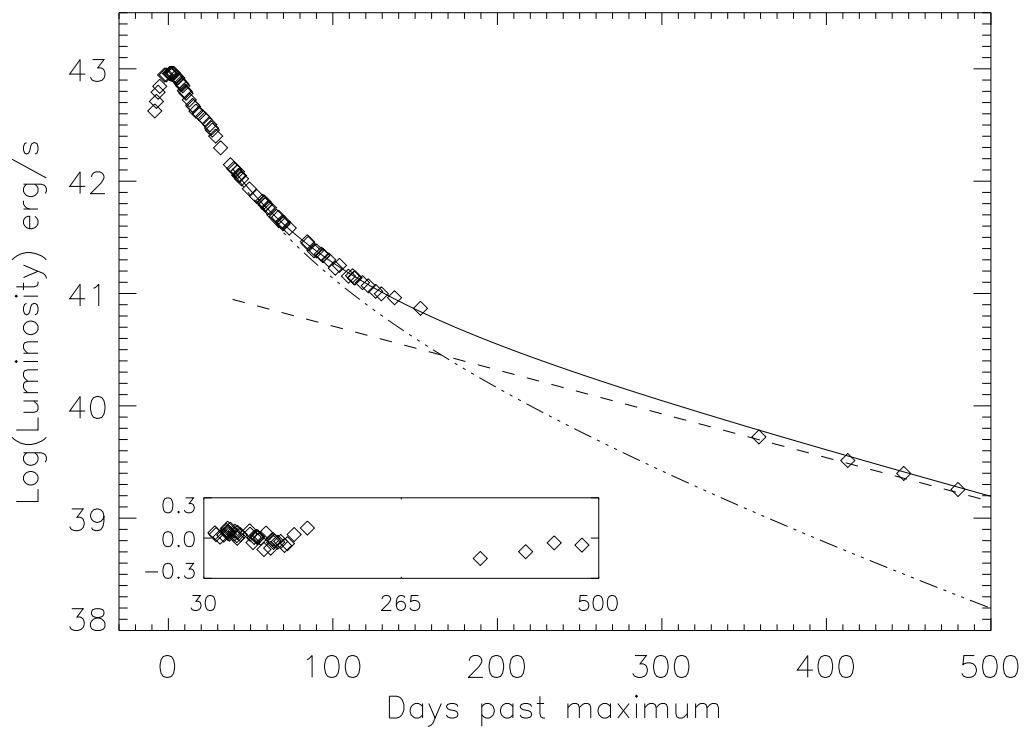}
\caption{Illustrative toy model of the bolometric luminosity for a 
SN~Ia light curve powered by 
radioactivity. Diamonds show our UVOIR light curve for SN 2000cx. 
The full line is the total luminosity for a simple $^{56}$Co decay 
model (see text). Assuming 0.28~\msun~of $^{56}$Ni,
the dot-dashed line shows the contribution from the $\gamma$-rays and the 
dashed line the contribution from the positrons. 
The inset shows the deviation in magnitudes versus time 
between the toy model and the data. The standard deviation is 0.05 mag.}
\label{toymodel}
\end{figure*}

\begin{figure*}[t]
\includegraphics[width=150mm,clip]{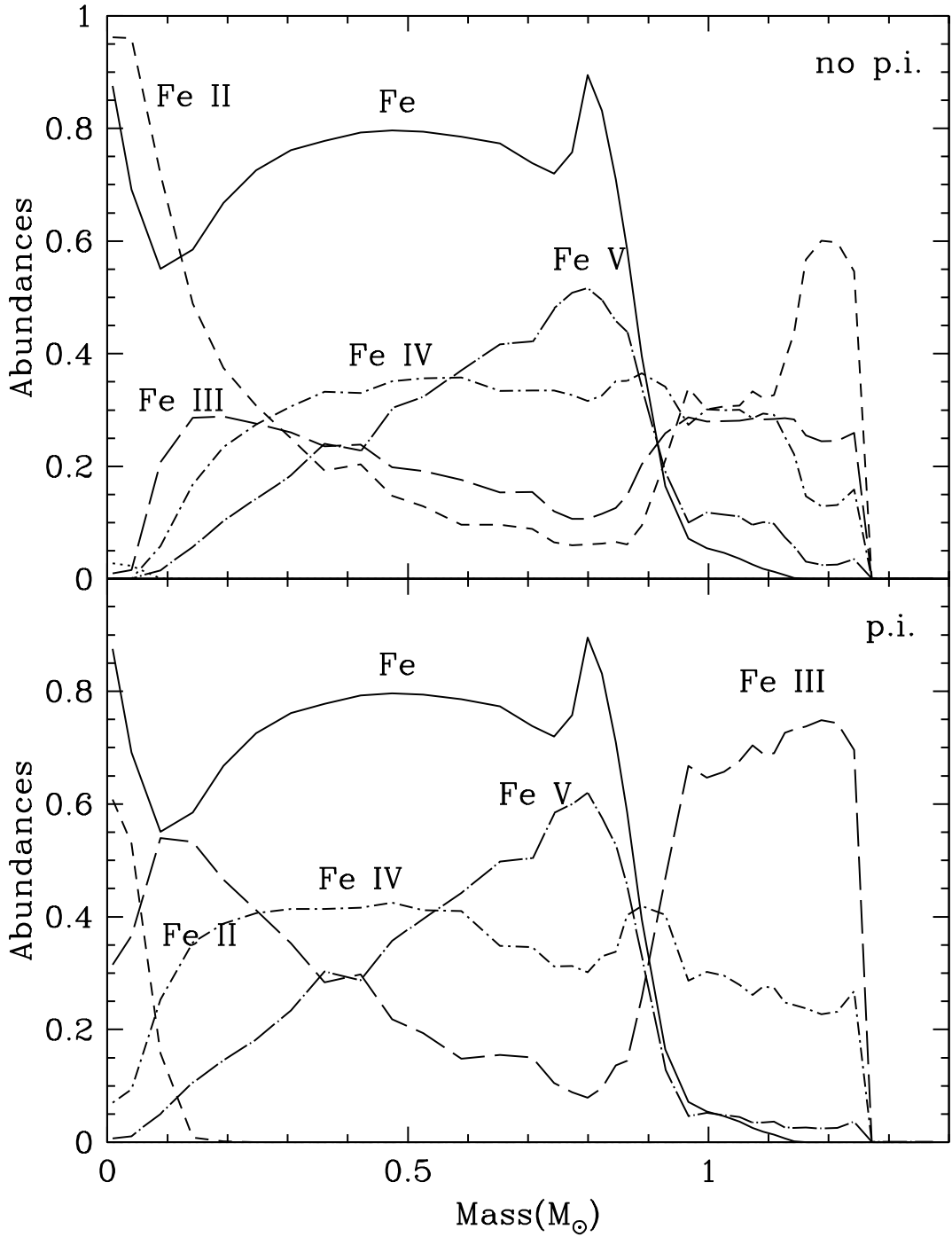}
\caption{
Calculated ionization structure of iron in the supernova ejecta at
300 days for the two different models.
The ionization for iron as a function
of mass coordinate in the ejecta. The upper figure shows 
the result for the model without photoionization and
the lower figure shows the result for the model including
photoionization. The solid curve shows the total iron
abundance. The broken curves show 
the fraction of the different iron ions.
}
\label{feion}
\end{figure*}

\begin{figure*}[t]
\includegraphics[width=140mm,clip,angle=-90]{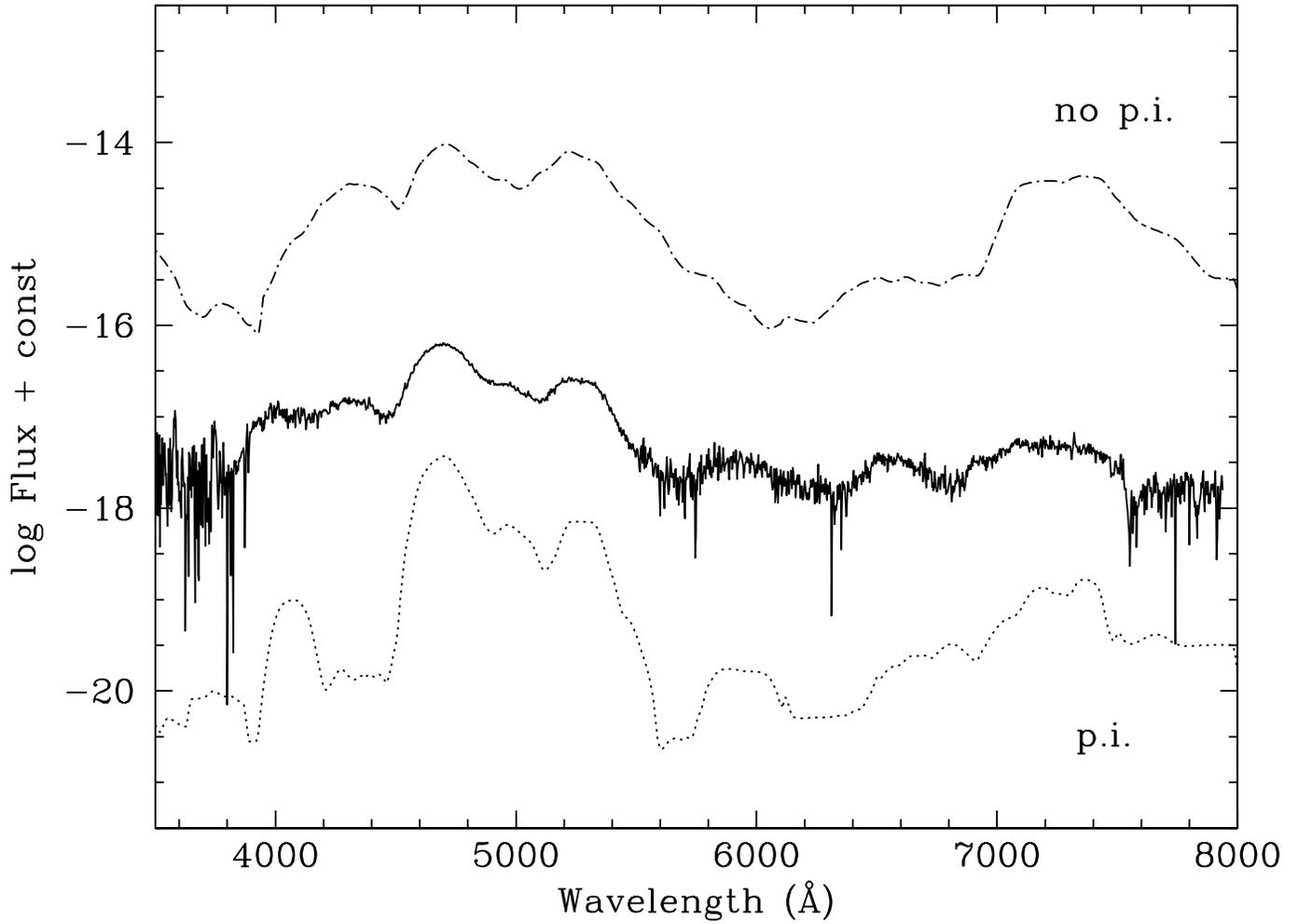}
\caption{Observed ({\it middle}) and modeled spectra of SN 2000cx at 
$\sim 360$ days. The {\it upper}
(dash-dotted) spectrum is a model calculation without photoionization, 
while the {\it lower} (dotted) spectrum is from a model including 
photoionization. The model spectra have been displaced by $\pm2$ dex, 
respectively.
The most prominent feature is the
emission between 4500 and 5500~\AA. In the model including photoionization
it is totally dominated by Fe~III emission, while 
in the model without photoionization 
it is a mixture 
of Fe~II and Fe~III emission. The feature at 4700~\AA\ is in this model 
mostly Fe~III,
while the emission at 5300~\AA\ is dominated by Fe~II.
From the comparison between observed and modeled spectra we find
somewhat better agreement for the model without photoionization. 
}
\label{spec}
\end{figure*}

\begin{figure*}[t]
\includegraphics[width=150mm,clip]{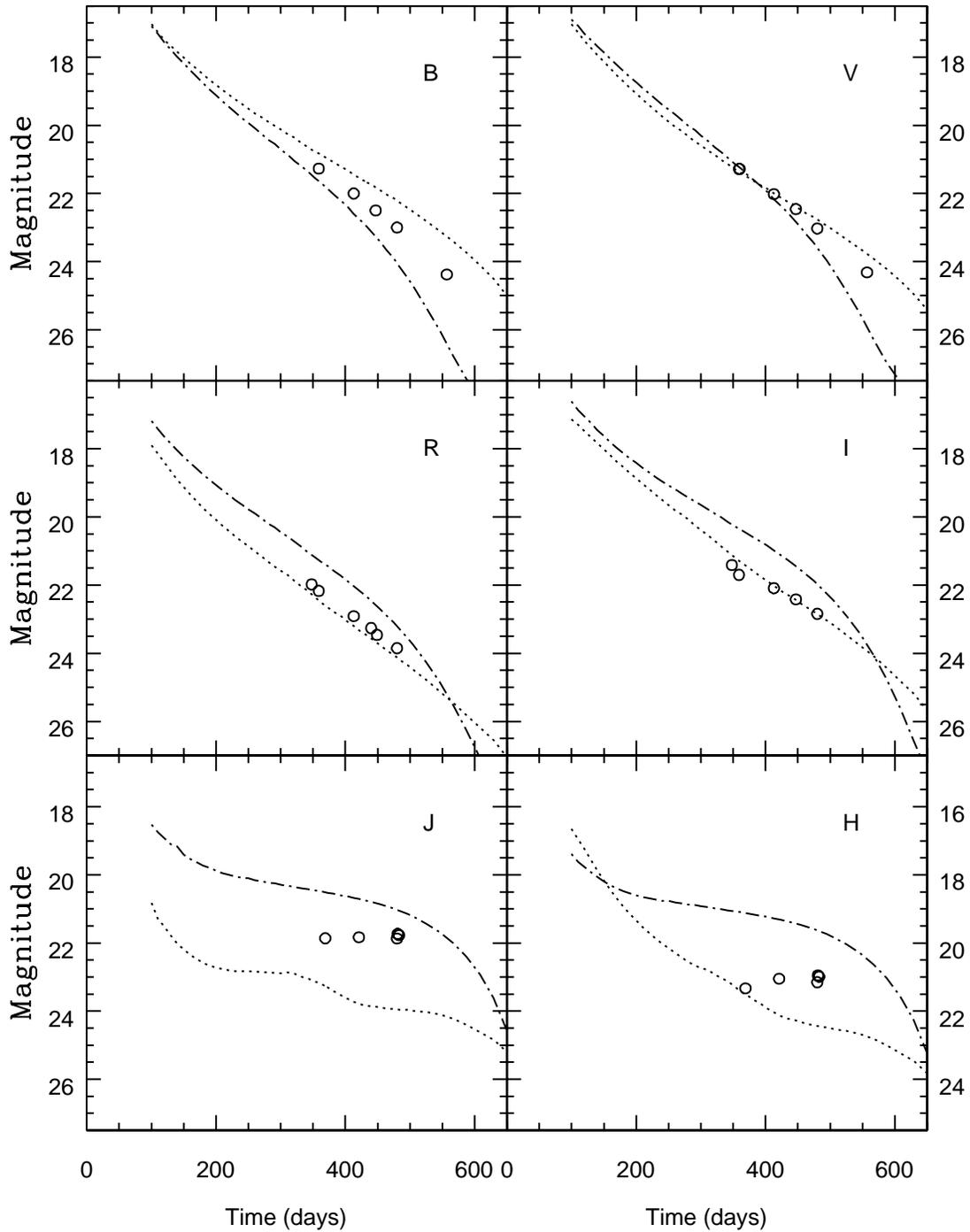}
\caption{Light curves for the $B$ to $H$ bands. The open circles are our
observations obtained with the VLT and {\it HST}. Two models are shown, 
one including
photoionization (dotted curve), and the other without photoionization 
(dash-dotted curve).
}
\label{phot}
\end{figure*}

\begin{figure*}[t]
\includegraphics[width=150mm,clip]{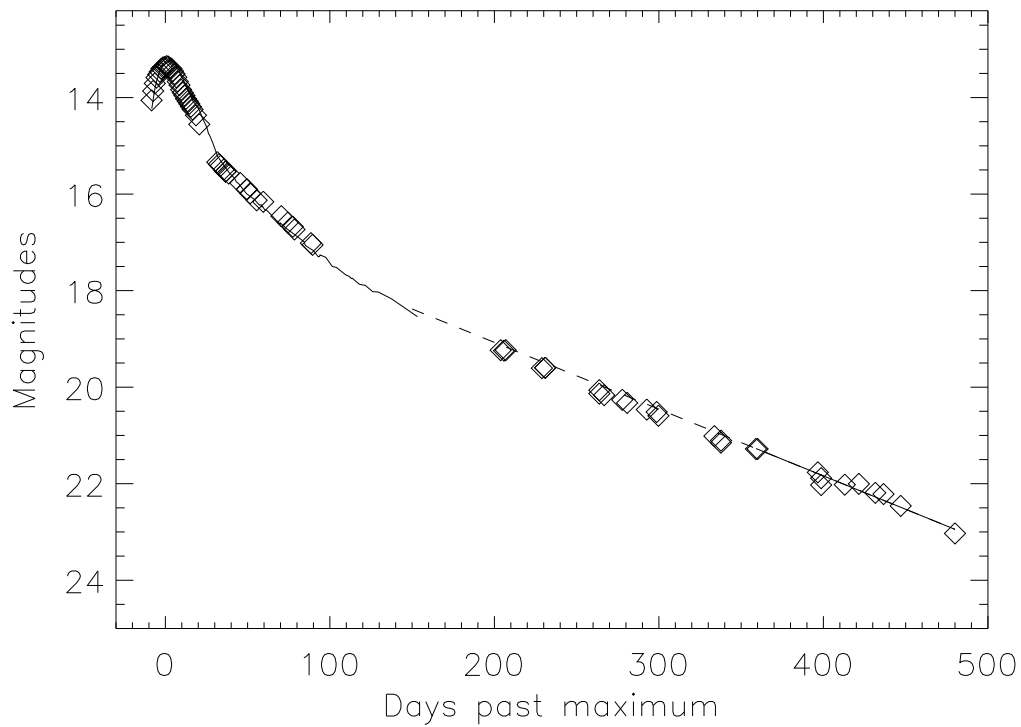}
\caption{SN 2000cx in the $V$ band (full line, see also Fig.~4) 
compared to SN 1992A (diamonds; Suntzeff 1996).
The magnitude and date of maximum for SN 1992A have been matched at peak to
the light curve of SN 2000cx. There is no indication that SN 2000cx behaves 
in a different way than SN 1992A at late phases.
SN 1992A had a very 
normal late-time light curve (e.g., Fig.~2 in Leibundgut 2000).}
\label{92a}
\end{figure*}

\begin{figure*}[t]
\includegraphics[width=150mm,clip]{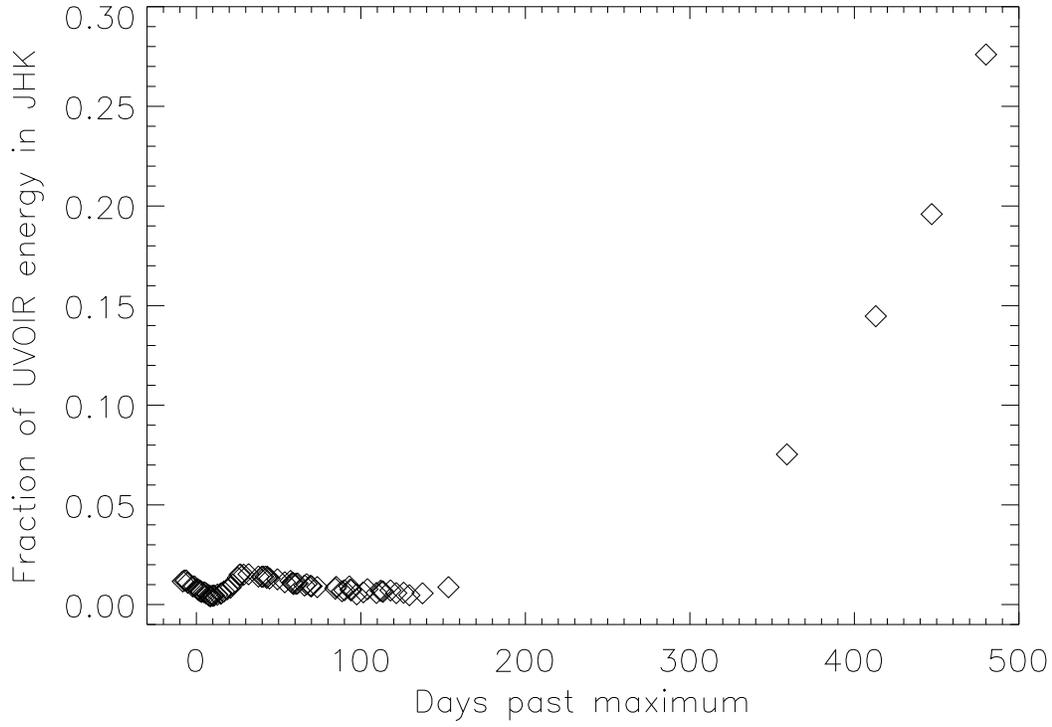}
\caption{Fraction of the luminosity in the near-IR bands as a 
function of time. 
The fraction is the luminosity in the near-IR bands divided with the  
total UVOIR luminosity
($L_{JHK}/L_{UBVRIJHK}$), where the UVOIR light curve 
has been constructed as described in the text.
}
\label{fraction}
\end{figure*}

\begin{figure*}[t]
\includegraphics[width=150mm,clip]{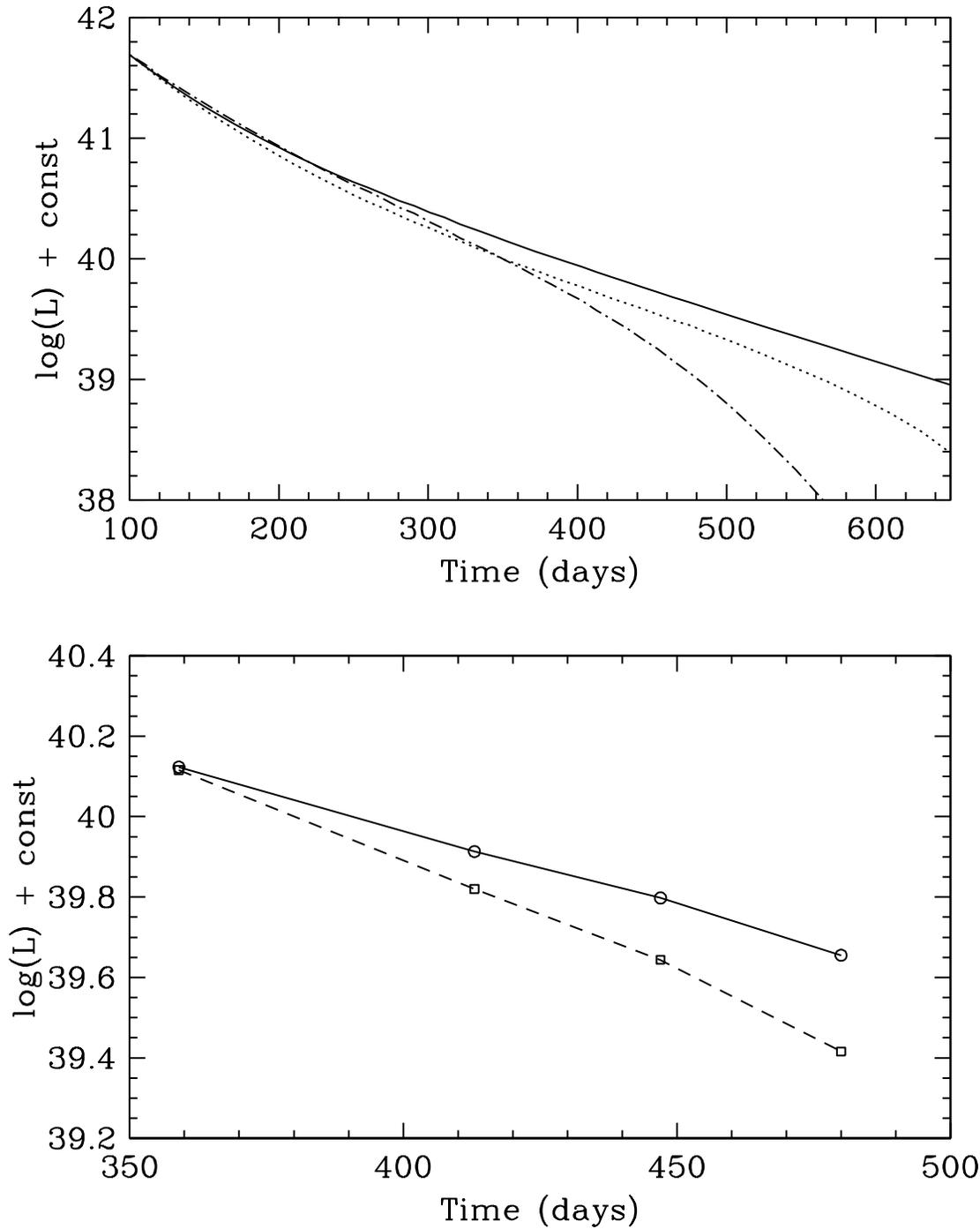}
\caption{
 Calculated (upper panel) and observed (lower panel) light curves
for SN 2000cx. The solid line in the upper panel shows the bolometric
luminosity, and the other lines show $V$-band luminosities 
for our two models.
The dotted curve refers to the model including photoionization, and
the dash-dotted curve refers to the model without photoionization.
The lower panel shows results from the observations. The solid curve
is the integrated luminosity in the $UBVRIJHK$ bands, while the dashed curve 
is the luminosity in the $V$ band. 
}
\label{bolv}
\end{figure*}

\end{document}